\documentclass[11pt,a4paper]{article}
\usepackage{jstyle}
\usepackage{tikz}
\usepackage{amsthm,mathrsfs,stmaryrd,bm, color}
\usepackage[vcentermath]{youngtab}
\usepackage{simplewick}
\usepackage{ulem}

\def\a{\alpha}
\def\b{\beta}
\def\g{\gamma}

\def\d{\delta}

\def\e{\epsilon}

\def\k{\kappa}
\def\l{\lambda}
\def\L{\Lambda}
\def\m{\mu}
\def\n{\nu}

\def\r{\rho}
\def\s{\sigma}

\def\t{\tau}
\def\f{\phi}

\def\vf{\varphi}

\def\o{\omega}

\author[a]{Euihun JOUNG}
\author[b]{\quad Karapet MKRTCHYAN}
\author[c]{\quad  Gabriel POGHOSYAN}
\affiliation[a]{Department of Physics and Research Institute of Basic
  Science, Kyung Hee University, Seoul 02447, Korea}
\affiliation[b]{Max Planck Institute for Gravitational Physics (Albert Einstein Institute)\\
Am M\"uhlenberg 1, 14476 Potsdam, Germany}
\affiliation[c]{Yerevan Physics Institute\\ Alikhanyan Br. Str. 2, 0036 Yerevan, Armenia}

\emailAdd{euihun.joung@khu.ac.kr}
\emailAdd{karapet.mkrtchyan@aei.mpg.de}
\emailAdd{gabrielpoghos@gmail.com}

\title{\centering
Looking for
partially-massless gravity}

\abstract{
We study the possibility for a unitary theory of partially-massless (PM) spin-two field interacting with Gravity in arbitrary dimensions. 
We show that the gauge and parity invariant interaction of PM spin two particles requires the inclusion of specific massive spin-two fields 
and leads to a reconstruction of Conformal Gravity, or multiple copies of the latter in even dimensions.
By relaxing the parity invariance, we find a possibility of a unitary theory in four dimensions, but this theory cannot be constructed in 
the standard formulation, due to the absence of the parity-odd cubic vertex therein.
Finally, by relaxing the general covariance, we show that a ``non-geometric'' coupling between massless and PM spin-two fields may lead to an alternative possibility of a unitary theory. 
We also clarify some aspects of interactions between massless, partially-massless and massive fields, and resolve disagreements in the literature.
}

\begin{document}

\maketitle

\section{Introduction}

One often encounters various obstructions when trying to explore the world of physical theories beyond the realm of ``standard models'' of gravity, gauge theories and matter.
Like the famous example of Coleman-Mandula theorem bypassed by supersymmetry, no-go theorems reveal their weak points 
when the key assumptions that they critically rely on are pinpointed.
In this paper, we examine such weak points of a specific no-go theorem, which dooms many 
possibly interesting extensions of gravity, through a relatively simple example --- the partially-massless (PM) gravity\footnote{We will call PM gravity any unitary theory of gravity that involves PM spin-two field.}.
Before introducing and reviewing the problem we want to address in PM gravity,
let us first comment on the generality of this no-go theorem, two classes of examples prohibited by it and weak points of the argument.

\paragraph{Admissibility condition}

The gauge invariance of a classical field theory has far-reaching implications.
As shown in \cite{Berends:1984rq,Barnich:1993vg}, general gauge symmetries are not restricted to the standard affine transformations
but allow dependence of fields in higher order. 
The invariance can be secured by a set of precise relations between the higher order parts 
of the gauge transformation and the Lagrangian.
Another consequence of gauge invariance is the closure of gauge transformation under commutator.
This also provides a set of non-trivial relations, which are simpler to implement in practice than the gauge invariance relations.
Especially focussing on the large gauge transformations with parameters satisfying the Killing tensor equations, namely the global symmetries,
one can easily arrive (see, e.g. \cite{Vasiliev:2004cm} and \cite{Joung:2014qya}) to the following two conclusions: first, the global symmetries defined as above should form a Lie algebra 
under the bracket defined through this procedure; second, the linearised on-shell fields should carry a representation
of such a global symmetry.
This second condition is what we refer to as {\it admissibility condition},
following the nomenclature of \cite{Konshtein:1988yg}\footnote{In \cite{Konshtein:1988yg}, the authors require unitarity of the corresponding representation. We will use a looser form of admissibility condition, allowing non-unitary theories in general, even though our eventual goal is to understand the possibility of unitary theories. We will show, that even without the requirement of unitarity, the admissibility condition puts very strong restrictions on the space of possible theories.}.

\paragraph{Two classes of examples} 

The admissibility condition is very powerful and often sufficient to rule out many illusive theories.
Here, let us mention two classes of examples, which were the main motivations for revisiting the PM gravity
as a toy model, which is of course interesting on its own.
The first class is the higher-spin theories  around (A)dS$_5$ background with  the global symmetries 
$\mathfrak{sl}_{\frac{N(N+1)(N+2)}{6}}$ (more precisely a certain real form of the latter complex algebra)  \cite{Fradkin:1989yd,Manvelyan:2013oua} (see also \cite{Boulanger:2011se}).
The above can be considered as a decent candidate for the global symmetry of a higher-spin gauge theory:
it contains the isometry algebra $\mathfrak{sl}_4\simeq \mathfrak{so}_6$ as a subalgebra
and additional generators corresponding to the Killing tensors 
of massless fields of spin $3,\ldots, N$.
Since there are only finite number of higher spin fields, 
these algebras seem to promise very simple and interesting models for higher spin gravity.
However, the admissibility condition does not allow a theory of massless fields of spin $2,\ldots, N$ 
because the corresponding on-shell field space --- or Hilbert space --- cannot carry 
a representation of (a real form of) $\mathfrak{sl}_{\frac{N(N+1)(N+2)}{6}}$.
Postponing the justification of why to the next paragraph, let us introduce the second class of  examples:
the PM higher-spin theories around  (A)dS$_{d+1}$ background \cite{Bekaert:2013zya,Grigoriev:2014kpa,Brust:2016zns} with the global symmetries $\mathfrak{sl}_{\frac{(d+1)_{k-1}(d+2k)}{k!}}$ \cite{Joung:2015jza}.
Again, it contains the isometry algebra $\mathfrak{so}_{d+2}$ as a subalgebra
 and additional generators corresponding to the Killing tensors of partially-massless fields of spin $s$ and depth $2t+1$\footnote{By this definition, the depth is the number of derivatives in the gauge transformation of PM field, which is different from \cite{Joung:2015jza}.}
 (hence, $0\le 2t\le s-1$)  with
 the relation $s-t\le k$. It is easy to check that the maximum spin is $2k+1$\,.
 Again these theories cannot satisfy the admissibility condition if we restrict the field content to the (partially-)massless fields
 corresponding to the symmetry generators. 
 We would like to remind here that the real forms of interest for the algebras in consideration here are non-compact, as they contain the isometry algebra of $(A)dS$ space, and therefore do not admit finite-dimensional unitary representations.

\paragraph{GK dimension analysis}

There is a simple and efficient way to check whether a certain infinite dimensional vector space is
large enough to carry a representation of a given Lie algebra.
The appropriate concept that
can be used to assess the ``size'' of an infinite dimensional representation
is so-called Gelfand-Kirillov (GK) dimension \cite{GK}.
Its proper definition is rather formal\footnote{See, e.g., $\href{https://en.wikipedia.org/wiki/Gelfand\%E2\%80\%93Kirillov_dimension}{\text{Wikipedia: Gelfand-Kirillov dimension}}$.}, and for applications in physics, it would be enough to regard it as the number of continuous variables required for a given representation to be realized as a space of functions of these variables (see, e.g., \cite{Joung:2014qya}). 
With this concept, it is simple to see that the GK dimension of a tensor sum representation is the larger one among 
the GK dimensions of two representations, and the GK dimension of a tensor product representation is the sum of the two
GK dimensions. 
Usual one-particle states in $D$ dimension have the Hilbert spaces with GK dimension $D-1$,
and the Hilbert space of finite number of such particles is still of GK dimension $D-1$.
Therefore, if the theories considered in the previous examples exist, then the GK dimensions of their Hilbert spaces are respectively 4 and $d$
because their field content is finite.
However, the Lie algebras $\mathfrak{sl}_{\frac{N(N+1)(N+2)}{6}}$ and $\mathfrak{sl}_{\frac{(d+1)_{k-1}(d+2k)}{k!}}$
do not admit representations\footnote{This does not rule out non-linear realisations of  global symmetries. An interesting recent example was provided in \cite{Bonifacio:2018zex}, a special Galileon theory in $(A)dS_{d+1}$ with extended symmetry (a real form of) $\mathfrak{sl}_{d+2}$, which is the $k=1$ case of the PM HS algebras of \cite{Joung:2015jza} mentioned above. The no-go statement here refers to the possibility of gauging these algebras and realising their global action linearly on the fields.} 
with such small GK dimensions for $N\ge3$ and $k\ge 2$.
In fact, all the infinite dimensional representations of $\mathfrak{sl}_{n}$ have GK dimensions not smaller than  $n-1$.
Therefore, this simple dimensional analysis rules out 
the theories based on these global symmetries. 

\paragraph{Possible bypass}

One of the crucial assumption in the above consideration is that the field content of the theories with the global symmetries
$\mathfrak{sl}_{\frac{N(N+1)(N+2)}{6}}$ or $\mathfrak{sl}_{\frac{(d+1)_{k-1}(d+2k)}{k!}}$
is composed of only gauge fields whose Killing tensors correspond to the generators of the symmetries.
In this way, the finiteness of the dimension of the symmetry algebras implies 
the finiteness of the number of fields in the theories.  
In fact, there is no strong reason that there should be only gauge fields. Indeed, a good example is Vasiliev's higher-spin gravity: 
the theory also requires 
a scalar field in the field content, which has no gauge parameter by itself.
Therefore, we may possibly bypass the obstruction imposed by the admissibility condition
by constructing a new field content which includes
an infinite number of additional non-gauge fields besides the finite number of the original gauge fields. 
In other words, we can induce a faithful representation of the global symmetry algebra
starting from the representation of the isometry subalgebra composed of on-shell gauge fields 
by adding additional vectors, that is, other fields.
This has a clear group theoretical meaning of ``induced representation'', hence the task could be worked out 
in the standard framework of the representation theory.
But, since our aim is to construct a classical Lagrangian for such a theory, 
it would be more useful to study the same problem in a field theoretical set-up.
In the two classes of theories mentioned above, it is likely that we need to deal with infinitely many fields of infinitely high spins and masses. 
The PM gravity, which we shall study in this paper, suffers from the same problem of validating the admissibility condition
but has a much simpler structure. 
Therefore, it can be a good starting example to study possible resolutions of the no-go theorem
given  by the admissibility condition.

\paragraph{Coming back to PM gravity}

In de Sitter space, there is a mass gap in the unitary spectrum of a spin-two particle, as opposed to 
the flat and anti-de Sitter spaces.
The lower mass bound of the massive spin-two particle is not the massless graviton, but
a very special massive particle, referred to as ``partially massless'' \cite{Deser:1983mm,Deser:2001pe,Deser:2001us,Deser:2001wx,Zinoviev:2001dt}. This mass value is also known as ``Higuchi bound'' \cite{Higuchi:1986py,Higuchi:1986wu,Higuchi:1989gz}. In four dimensions, the partially-massless spin-two (PM) field has 4 degrees of freedom (DoF) -- more than those of massless field (2 DoF) and less than those of massive field (5 DoF). 
The difference from the massive DoF is the scalar mode, which
invokes various problems in the consistency of massive gravity.
Recently, the PM fields attracted a lot of attention \cite{Metsaev:2006zy,Skvortsov:2006at,Francia:2008hd,Metsaev:2009hp,Alkalaev:2011zv,Afshar:2011yh,Afshar:2011qw,
Joung:2012rv,Deser:2012qg,Hassan:2012gz,Joung:2012hz,Deser:2013uy,deRham:2013wv,Hassan:2013pca,Brito:2013yxa,Zinoviev:2014zka,Joung:2014aba,Hinterbichler:2014xga,Gover:2014vxa, Alexandrov:2014oda,Garcia-Saenz:2014cwa, Hinterbichler:2015nua,Hassan:2015tba,Cherney:2015jxp,Garcia-Saenz:2015mqi,
Gwak:2016sma,Hinterbichler:2016fgl,Apolo:2016ort,Apolo:2016vkn,Bonifacio:2016blz,
Bernard:2017tcg,Basile:2017kaz,Galviz:2017tda,Boulanger:2018shp,Zinoviev:2018eok,Tannukij:2018vqm,Galviz:2018eep,Garcia-Saenz:2018wnw,Fortes:2019asi,Buchbinder:2019olk}
partly due to their potential relevance in the cosmological phenomena \cite{Baumann:2017jvh,Goon:2018fyu},
and there have been various attempts to construct a consistent gravity theory of PM field with/without massless graviton.
However, it turns out that such a theory is prohibited by several no-go results\footnote{For the moment, 
the only potential exceptions require exotic set-ups like non-local field theory \cite{Hassan:2013pca,Hassan:2015tba,Tannukij:2018vqm}.} \cite{deRham:2013wv,Deser:2013uy,Hassan:2012gz,Joung:2014aba,Apolo:2016vkn,Garcia-Saenz:2015mqi,Apolo:2016ort,Zinoviev:2018eok}.
Here, we revisit the no-go theorem \cite{Joung:2014aba}:
an interacting theory involving massless and PM spin two fields in a four dimensional dS background 
has the $\mathfrak{so}(1,5)$ as its global symmetry algebra,
but the analysis of the admissibility condition shows that the same field content
cannot form a representation of $\mathfrak{so}(1,5)$.
This obstruction can be avoided if the massless spin two
and PM spin two fields have relatively negative kinetic term signs.
In such a case, the global symmetry becomes $\mathfrak{so}(2,4)$ and in fact the resulting theory is nothing but
the conformal gravity written in two-derivative form around a constant curvature background \cite{Deser:2012qg}.
In the following sections, we shall attempt to fix the problem of PM gravity with 
the $\mathfrak{so}(1,5)$ (or, as we show, $\mathfrak{so}(1,D+1)$ in arbitrary dimensions $D\geq 4$) global symmetry by enlarging the field content
with additional fields or relaxing implicit assumptions of the no-go theorem, such as parity invariance and general covariance.

\paragraph{Organization of the paper}
\begin{itemize}
\item
In Section \ref{sec: gauge invariance},
we shall briefly review how the general gauge invariance
gives rise to a set of conditions --- global symmetry condition and admissibility condition --- which shall be used in the later analysis.
\item
In Section \ref{sec: PM no-go},
we apply such conditions to the theory of interacting massless and PM spin two fields
and show how the global symmetry $\mathfrak{so}(1,D+1)$ arises and how the admissibility condition,
that is the closure of the global symmetry on each fields, is violated 
within the setting of massless and PM spin two fields.
The content up to this point is basically a summary of \cite{Joung:2014aba}, generalised to arbitrary dimensions.
\item
In Section \ref{sec: enlarge FC},
we discuss a possible remedy of the problem by enlarging the field content of the theory.
We show that only the addition of massive spin-two fields can fix the non-closure problem on the PM field,
but then the same problem reappears for the newly introduced field. 
This requires an iterative introduction of more and more massive spin two fields. 
We find that the non-closure problem becomes  incurable after a few iterations when the mass of added spin two field reaches a certain bound.
This problem is very much analogous to the original problem that we began with:
the admissibility condition imposes a certain algebraic equation for the coupling constants
which have non-trivial solution only when the relative kinetic term sign of massless and PM spin two fields is negative.
This non-unitary resolution is also possible in any even $D$, but
not in odd dimensions $D\geq 5$.
In even dimensions, we find that the minimal possibility is Conformal Gravity, while
alternatives are related to the multiples of the field content of Conformal Gravity.
We discuss various aspects of these putative theories.

\item
In Section \ref{sec: other assum}, we discuss a possible remedy of the problem by relaxing other assumptions.
First, we relax the parity invariance
and show that the $\mathfrak{so}(1,5)$ transformation can be realized in terms of two fields, the massless and PM spin two fields, in four dimensions. However, the corresponding Lagrangian cubic vertex is obstructed in the 
formulation of covariant metric-like fields.
Second, we relax the condition of the general covariance allowing for non-generally-covariant (or, ``non-geometric'')
interactions between massless and (partially) massive spin two fields. 
This lets us identify the problematic non-closure part with
the modification of the transformation rule induced by the addition of the non-generally-covariant interaction vertex.
However, this modification invokes the problem of symmetry non-closure in other commutators.
This can be potentially fixed by introducing yet another field of fourth rank. The irreducible components of the new field are fully antisymmetric and ``window'' Young-diagram type.

\item
In Section \ref{sec: discussion},
we summarize the content of the paper and 
discuss open directions.

\item
Appendix \ref{sec: details} contains various technical details. Appendix \ref{sec: Cubic vertices} contains details on cubic vertices with massless, PM and massive spin-two fields. Here we also discuss subtle aspects of field redefinition freedom and gauge transformation deformations due to interactions involving (partially) massive fields.
In Appendix
\ref{sec: scalar coupling}
we study the PM coupling to matter. We show there that the mass value of matter is constrained by the requirement that it couples to PM field.
\end{itemize}

\section{Consequences of Gauge Invariance}
\label{sec: gauge invariance}

Let us consider a generic theory with a  field content $\{\chi_i\}$ and an action $S$, which can be perturbatively expanded around some background field configuration, starting from linearised quadratic action that is diagonal with respect to all the fields in the theory. 
The gauge invariance, $\delta_\varepsilon S=0$\,, implies 
the closure of the gauge symmetry {(see, e.g., \cite{Berends:1984rq})}:
\be
	\delta_{\varepsilon}\,\delta_{\eta}-\delta_{\eta}\,\delta_{\varepsilon}
	=\delta_{[\eta,\varepsilon]}
	+C_{ij}(\eta,\varepsilon)\,\frac{\delta S}{\delta \chi_i}\,\frac{\d}{\d\chi_j}\,.
	\label{closure}
\ee
Here $[\eta,\varepsilon]$ and $C_{ij}$ are a priori field-dependent and $C_{ij}$ is antisymmetric under the exchange of $i$ and $j$. 
The relation \eqref{closure} is easy to motivate. The left hand side, acting on the action, should give zero due to gauge invariance of the action. Therefore, the right hand side should be a gauge transformation, up to a trivial transformation.
Expanding all the field-dependent quantities in the power of fields as
\ba
	&S=S^{\sst [2]}+S^{\sst [3]}+\cdots, \qquad & \delta_\e=\delta_\e^{\sst [0]}+\delta_\e^{\sst [1]}+\cdots,\nn
	&[\eta,\varepsilon]=[\eta,\varepsilon]^{\sst (0)}+[\eta,\varepsilon]^{\sst (1)}+\cdots, 
	\qquad & C_{ij}=C_{ij}^{\sst (0)}+C_{ij}^{\sst (1)}+\cdots,
\ea
we can derive several conditions which we shall use in the following analysis:
\begin{enumerate}
\item From the next-to-lowest gauge invariance condition,
\be
	\delta^{\sst [0]}S^{\sst [3]}+\delta^{\sst [1]}S^{\sst [2]}=0\,,
\ee
 we can find all possible cubic vertices and the corresponding first order gauge transformations, 
 \be
 	S^{\sst [3]}(\chi_i, \chi_j, \chi_k)\quad \longrightarrow \quad 
	\left[\delta^{\sst [1]}_{\e_i}\chi_j\right]_{\chi_k}\ \textrm{and the permutations of $i, j, k$}.
\ee
Here $[\,\cdots]_\chi$ means the part linear in $\chi$.
It is important to note that in this way we find the linear part of gauge transformation
--- which will give the global symmetry transformation later ---
projected to the pre-assumed field content.
Therefore, by including an additional field, say $\phi$, we can have additional cubic vertices $S^{\sst [3]}(\chi_i,\chi_j, \phi)$
and the transformation of the original field $\delta^{\sst [1]}_{\e_i}\chi_j$ can acquire
an additional term $\left[\delta^{\sst [1]}_{\e_i}\chi_j\right]_\phi$.

\item With $\delta^{\sst [1]}$ derived in this way, we can then calculate the lowest order of the commutator 
$[\cdot,\cdot]^{\sst [0]}$ from the lowest order part of the closure condition \eqref{closure} as
\be
	\delta^{\sst [0]}_{\varepsilon}\,\delta^{\sst [1]}_{\eta}
	-\delta^{\sst [0]}_{\eta}\,\delta^{\sst [1]}_{\varepsilon}=\delta^{\sst [0]}_{[\eta,\varepsilon]^{\sst [0]}}\,.
	\label{cond 1}
\ee

\item Focussing on the  Killing tensors $\bar\varepsilon$ satisfying 
\be
	\delta^{\sst [0]}_{\bar\varepsilon}=0\,,
\ee
one can verify that the condition \eqref{cond 1} gives
\be
	\delta^{\sst [0]}_{[\bar\eta,\bar\varepsilon]^{\sst [0]}}=0\,,
\ee
that is to say that the global symmetry is closed under $[\bar\eta,\bar\varepsilon]^{\sst [0]}$.
Hence, we can take it as the Lie bracket of this Lie algebra:
$[\![ \bar \eta,\bar\varepsilon]\!]:=[\bar\eta,\bar\varepsilon]^{\sst [0]}$.
We can also require at this stage that the Lie bracket
defined in this way does satisfy Jacobi identity.

\item Moving to the next-to-lowest part of the closure condition \eqref{closure} with the Killing tensors, we find 
\be
	\delta^{\sst [1]}_{\bar\varepsilon}\,\delta^{\sst [1]}_{\bar\eta}
	-\delta^{\sst [1]}_{\bar\eta}\,\delta^{\sst [1]}_{\bar\varepsilon}
	=\delta^{\sst [1]}_{[\![\bar\eta,\bar\varepsilon]\!]}
	+\delta^{\sst [0]}_{[\bar\eta,\bar\varepsilon]^{\sst [1]}}
	+C^{\sst [0]}_{ij}(\bar\eta,\bar\varepsilon)\,\frac{\delta S^{\sst [2]}}{\delta \chi_i}\,\frac{\d}{\d\chi_j}\,.
	\label{admin 1}
\ee

By applying the above condition to a free one-shell field $\chi_i$, we get 
\be
	\left(\delta^{\sst [1]}_{\bar\varepsilon}\,\delta^{\sst [1]}_{\bar\eta}
	-\delta^{\sst [1]}_{\bar\eta}\,\delta^{\sst [1]}_{\bar\varepsilon}\right)\chi_i
	\sim \delta^{\sst [1]}_{[\![\bar\eta,\bar\varepsilon]\!]}\,\chi_i\,.
	\label{admin cond}
\ee

Note that the last term of \eqref{admin 1} vanishes due to the free on-shell equation $\delta S^{\sst [2]}/\delta\chi_i=0$,
and the next-to-last term is equivalent to zero where
the equivalence relation $\sim$ is the usual one of the free on-shell field,
\be
	\chi_i \sim \chi_i+\delta^{\sst [0]}_{\varepsilon} \chi_i\,.
\ee
Or, equivalently we can apply \eqref{admin 1} to the quadratic action $S^{\sst [2]}$ ending up with
\be
	\left(\delta^{\sst [1]}_{\bar\varepsilon}\,\delta^{\sst [1]}_{\bar\eta}
	-\delta^{\sst [1]}_{\bar\eta}\,\delta^{\sst [1]}_{\bar\varepsilon}\right)S^{\sst [2]}	
	= \delta^{\sst [1]}_{[\![\bar\eta,\bar\varepsilon]\!]}\,S^{\sst [2]}\,.
	\label{admin equiv}
\ee
The condition \eqref{admin cond} or \eqref{admin equiv} is what we refer to as {\it admissibility condition}.
It is nothing but the condition that the space of on-shell fields carries 
a representation of the global symmetry algebra.

\end{enumerate}

Note, that in all of this procedure only in the equation \eqref{admin cond} we take the fields to be on-shell, satisfying $\delta S^{\sst [2]}/\delta\chi_i=0$. This general procedure will be applied to the particular example in the following Sections. 

\section{Review: No-Go on PM gravity}
\label{sec: PM no-go}

In this Section, we shall apply the conditions derived in Section \ref{sec: gauge invariance}
to the setting of massless and PM spin two fields.
In \cite{Joung:2014aba}, the analysis has been carried out in a generic gravitational background
so that one could focus on the PM interaction part, as the general covariance would ensure that there are no problems with gravitational interaction.

Here, we shall make the set-up simpler by starting from the dS background
and by studying perturbative consistency of both massless and PM spin two fields.
The massless spin two interactions can be completed in a way reproducing Einstein's gravity,
while the PM interactions will be exposed to severe consistency examinations, generalising the analysis of \cite{Joung:2014aba} to arbitrary dimensions.

\paragraph{Quadratic theory}

The theory we shall examine contains two fields:
massless spin two field $h_{\mu\nu}$ and PM spin two field $\varphi_{\m\n}$.
Their free action, that is the quadratic part $S^{\sst [2]}[h,\varphi]$ of the conceivable full action $S[h,\varphi]$, 
is given by
\be
	S^{\sst [2]}[h,\varphi]
	=\frac12 \int d^D x\sqrt{-\bar g}
	\left[ h^{\m\n}\,\cG^{(0)}_{\m\n}(h) 
	+\varphi^{\m\n}\,\cG_{\m\n}^{(\frac{2\L}{D-1})}(\varphi)\right],
	\label{S2}
\ee 
where $\bar g_{\m\n}$ is the dS metric 
satisfying $\bar R_{\m\n,\r\s}=\frac{2\,\Lambda}{(D-1)(D-2)}\left(\bar g_{\m\r}\,\bar g_{\n\s}-\bar g_{\n\r}\,\bar g_{\m\s}\right)$,\footnote{Here, 
we use the convention $$[\nabla_\m, \nabla_\n]\,V^{\r}{}_{\l}=\bar R_{\m\n,\l}{}^{\s}\,V^{\r}{}_{\s}-
\bar R_{\m\n,\s}{}^{\r}\,V^{\s}{}_{\l}=
\frac{2\,\Lambda}{(D-1)(D-2)}\left(\bar g_{\m\l}\,V^\r{}_{\n}-\bar g_{\n\l}\,V^\r{}_\m
-\d^\r_\n\,V_{\m\l}+\d^\r_\m\,V_{\n\l}\right).$$}
and all the indices are raised and lowered with this background metric.
Massless and PM spin two fields correspond to the $m^2=0$ and $m^2=\frac{2\,\L}{D-1}\equiv \frac{D-2}{L^2}$ points
of the massive spin two theory,\footnote{We use the $dS$ radius $\frac{1}{L^2}=\frac{2\,\L}{(D-1)(D-2)}$. In the case of $AdS$, one should change $\frac1{L^2}\rightarrow -\frac1{L^2}$.} where the two derivative operator takes the form,
\ba
\mathcal{G}^{(m^2)}_{\m\n}(\phi)
\eq \left(\Box-\tfrac{2}{L^2}-m^2\right) \phi_{\m\n}-2\,\nabla_{(\m}\,\nabla^\rho\,\phi_{\n)\rho}
+\nabla_{\m}\,\nabla_{\n}\,\phi^\rho{}_\rho\nn
&&-\,\bar g_{\m\n} \left[ \left(\Box+\tfrac{D-3}{L^2}-m^2\right) \phi^\rho{}_\rho-\nabla^\rho\,\nabla^\sigma\,\phi_{\rho\sigma} \right],\label{Gm}
\ea
where the d'Alembertian operator $\Box$ and the covariant derivative $\nabla$ are also defined 
with the dS metric $\bar g_{\m\n}$.
The massless spin two action with $\cG_{\m\n}^{(0)}$ is nothing but the quadratic part of the Einstein-Hilbert action with cosmological constant, $\int d^D x \sqrt{-g}\,(R-2\L)$\,, expanded in the metric perturbation: $g_{\m\n}=\bar g_{\m\n}+h_{\m\n}$\,.

 The gauge symmetries of the quadratic action \eqref{S2} are
 \be
 \d^{\sst [0]}_\xi h_{\m\n}=\nabla_{(\m}\,\xi_{\n)}\,,\qquad
\d^{\sst [0]}_\a\varphi_{\m\n}=\left(\nabla_{\m}\,\partial_{\n}+\tfrac{1}{L^2}\,\bar g_{\m\n}\right)\a\,,
\ee
where the $h_{\m\n}$ transformation is just linearised diffeomorphism.
The PM spin-two field $\varphi_{\m\n}$ has a two-derivative gauge symmetry with scalar parameter $\a$. 
 
It would be useful to write here the set of on-shell conditions for these spin-two fields.
For a generic mass squared $m^2$, it is 
\be
	 \left(\Box-\tfrac{2}{L^2}-m^2\right) \phi_{\m\n}=0\,,
	 \quad 
	 \nabla^{\m}\,\phi_{\m\n}=0\,,
	 \qquad
	 \phi_{\m}{}^\m=0\,.
	 \label{OS}
\ee
The above mentioned two special cases are supplemented with gauge symmetries. The massless field $h_{\m\n}$ satisfies the conditions \eqref{OS} with $m^2=0$ together 
with the equivalence relation,
\be
	h_{\m\n}\sim h_{\m\n}+ \nabla_{(\m}\,\xi_{\n)}
	\qquad
	[\,\nabla^\m\xi_\m=0\,,\quad (\Box+ \tfrac{D-1}{L^2})\,\xi_\m=0\,]\,,
\ee
while the PM field satisfies the conditions \eqref{OS} with $m^2=\frac{2\L}3$ and the equivalence relation,
\be
	\varphi_{\m\n}\sim \varphi_{\m\n}+ \left(\nabla_{\m}\,\partial_{\n}+\tfrac{1}{L^2}\,\bar g_{\m\n}\right)\a
	\qquad
	[\,(\Box+\tfrac{D}{L^2})\,\a=0\,]\,.
	\label{OPM}
\ee

\paragraph{Cubic interactions and first order gauge transformations.}

We move to the cubic part of the action, $S^{\sst [3]}$,
and the corresponding $\delta^{\sst [1]}$.
Since we have two fields $h_{\m\n}$ and $\varphi_{\m\n}$,
there are four types of cubic vertices:
$h\!-\!h\!-\!h$, $h\!-\!h\!-\!\varphi$, $h\!-\!\varphi\!-\!\varphi$ and $\varphi\!-\!\varphi\!-\!\varphi$\,.
By focusing on the two-derivative interactions, 
 we find that $h\!-\!h\!-\!\varphi$ is not allowed whereas
 $\varphi\!-\!\varphi\!-\!\varphi$ is available only in four dimensions.
In principle, we can allow the interactions with more than two derivatives, but their presence does not 
affect the consistencies (global symmetry and admissibility condition) of two derivative couplings.
For this reason, we can disregard these higher derivative couplings at the moment
and consider  
\be
	S^{\sst [3]}[h,\varphi]
	=\int d^D x\, \sqrt{-\bar g}\,\Big(
	\l_{hhh}\,\cV_{hhh}(h,h,h)
	+\l_{h\varphi\varphi}\,\cV_{h\varphi\varphi}(h,\varphi,\varphi)
	+\l_{\varphi\varphi\varphi}\,\cV_{\varphi\varphi\varphi}(\varphi,\varphi,\varphi)\Big)\,,
	\label{V012}
\ee
where $\cV_{hhh}, \cV_{h\varphi\varphi}, \cV_{\varphi\varphi\varphi}$ are the couplings which contain at most two derivatives.
The vertices $\cV_{hhh}$ and $\cV_{h\varphi\varphi}$ are nothing but the ones appearing in the Einstein gravity
and the gravitational minimal coupling for the quadratic PM action.
Finally $\cV_{\varphi\varphi\varphi}$ is the vertex which exists only in four dimensions and can be extracted from the four dimensional conformal gravity.
In the current context, it is not important whether $\cV_{hhh}, \cV_{h\varphi\varphi}, \cV_{\varphi\varphi\varphi}$ arise in specific non-linear theories
but the fact that they form a basis for any gauge invariant cubic interaction with no more than two derivatives.\footnote{Note
 that in fact there exists one more $h\!-\!\varphi\!-\!\varphi$ two-derivative coupling which is independent from the gravitational one $\cV_{h\varphi\varphi}$. 
We also disregard this possibility here as it would spoil the general covariance of the PM field, 
but will come back to it in Section \ref{sec: non-grav}.
}
The precise expressions of $\cV_{hhh}, \cV_{h\varphi\varphi}, \cV_{\varphi\varphi\varphi}$ are not important here and can be found in Appendix \ref{sec: EH}.
It is also worth to note that up to this stage the coupling constants $\l_{hhh}, \l_{h\varphi\varphi}, \l_{\varphi\varphi\varphi}$ are arbitrary.
The first order part of the gauge transformation, $\delta^{\sst [1]}$, can be extracted from each of these cubic vertices by taking a gauge variation with respect to $h$ or $\varphi$. 
First, by taking $h$-variation, we obtain
\ba
 \delta^{\sst [1]}_{\xi}\,h_{\m\n}
 \eq \l_{hhh}\left(\xi^\rho\,\nabla_\rho\,h_{\m\n}
 +2\,\nabla_{(\m}\,\xi^\rho\,h_{\n)\rho}\right),
 \label{d1h}
 \\
  \delta^{\sst [1]}_{\xi}\,\varphi_{\m\n}
 \eq \l_{h\varphi\varphi}\left(\xi^\rho\,\nabla_\rho\,\varphi_{\m\n}
 +2\,\nabla_{(\m}\,\xi^\rho\,\varphi_{\n)\rho}\right),
 \label{d1phi}
\ea
where  $\xi^\mu=\bar g^{\m\n}\,\xi_\n$.
Second, by taking $\varphi$-variation, we obtain ($\l_{\vf\vf\vf}\neq 0$ only for $D=4$)
\ba
 \delta^{\sst [1]}_{\a}\,h_{\m\n}
 \eq 2\,\l_{h\varphi\varphi}\,\Big[\partial^\rho\a
 \left(\nabla_\rho\,\varphi_{\m\n}-2\,\nabla_{(\m}\,\varphi_{\n)\rho}\right)
 -\frac{D-4}{L^2}\,\a\,\vf_{\m\n}\Big],\nn
  \delta^{\sst [1]}_{\a}\,\varphi_{\m\n}
 \eq
\frac12\,\l_{h\varphi\varphi}\left[\partial^\rho\a
 \left(\nabla_\rho\,h_{\m\n}-2\,\nabla_{(\m}\,h_{\n)\rho}\right)
 +\frac{2}{L^2}\,\a\,h_{\m\n}\right]\nn
 &&+\,
2\,\l_{\varphi\varphi\varphi}\,\partial^\rho\a
 \left(\nabla_\rho\,\varphi_{\m\n}-\nabla_{(\m}\,\varphi_{\n)\rho}\right).
 \label{h varphi transf}
\ea
Then we can extract the zeroth order of the commutator as
\ba
	\,[\,\xi_1,\xi_2\,]^{\sst [0]} \eq \l_{hhh}\, \xi_{[1}^\mu\,\partial_\m\,\xi_{2]}^\n\,\partial_\n\,,\nn
	\,	[\,\xi,\a\,]^{\sst [0]} \eq \l_{h\varphi\varphi}\, \xi^\mu\,\partial_\m\,\a\,,\nn
	\,[\,\a_1,\a_2\,]^{\sst [0]}\eq -2\,\l_{h\varphi\varphi} \left(\partial_\rho\,\a_{[1}\,\nabla^\m\partial^\r\,\a_{2]}
	+\frac{D-4}{L^2}\,\a_{[1}\partial^\m\a_{2]}\right)\,\partial_\m\,.
	\label{commutator}
\ea
The first and third commutators are vectors, hence close as a $h$-gauge transformation (diffeomorphism).
The second commutator is a scalar and closes as a $\varphi$-gauge transformation.

\paragraph{Killing tensors and global symmetries} The solution space of the Killing equations,
\be
	\delta^{\sst [0]}_{\bar\xi}h_{\m\n}=\nabla_{(\m}\,\bar\xi_{\n)}=0\,,
	\qquad
	\d^{\sst [0]}_{\bar\a}\varphi_{\m\n}=\left(\nabla_{\m}\,\partial_{\n}+\tfrac{1}{L^2}\,\bar g_{\m\n}\right)\bar\a=0\,,
	\label{Killing}
\ee
are generated, respectively, by
\be
	M^{\sst AB}_\m=2\,L^2\,\frac{X^{\sst [A}\,\partial_\m\,X^{\sst B]}}{X^2}\,,
	\qquad 
	K^{\sst A}=L\,\frac{X^{\sst A}}{\sqrt{X^2}}\,,
\ee
where we used the ambient space formulation {(see e.g. \cite{Joung:2011ww} for the details of the ambient formulation)}: 
here dS$_4$ is described by the equation $X^2=L^2$\,.
If we compute the Lie bracket, $[\![\cdot,\cdot]\!]=[\cdot,\cdot]^{\sst [0]}$ 
{(where $[\cdot,\cdot]^{\sst [0]}$ is defined in \eqref{cond 1})}, between 
such generators of Killing tensors,
we find
\ba
	\big[\hspace{-3pt}\big[\,M^{\sst AB}\,,\,M^{\sst CD}\,\big]\hspace{-3pt}\big]
	\eq \l_{hhh}\left(\eta^{\sst AD} M^{\sst BC}+\eta^{\sst BC} M^{\sst AD} - \eta^{\sst AC}\,M^{\sst BD} - \eta^{\sst BD} M^{\sst AC}\right),
	\label{isometry}\\
	\big[\hspace{-3pt}\big[\,M^{\sst AB}\,,\,K^{\sst C}\,\big]\hspace{-3pt}\big] \eq 
	\l_{h\varphi\varphi}\left(\eta^{\sst BC}\,K^{\sst A} -\eta^{\sst AC}\,K^{\sst B}\right),
	\label{general cov}\\
\big[\hspace{-3pt}\big[\,K^{\sst A}\,,\,K^{\sst B}\,\big]\hspace{-3pt}\big] \eq  -\frac{D-3}{L^2}\,\l_{h\varphi\varphi}\,M^{\sst AB}\,.
	 \label{PM com}
\ea
{For more details of the derivation, we refer to \cite{Joung:2014aba} (section 3.2 of the arXiv version)
where the computation was carried out with $D=4$. In the ambient formulation the dimensionality enters only as the parameter $D$, so we can straightforwardly generalize the result of \cite{Joung:2014aba} to any $D$.}
By asking Jacobi identity to hold, we find 
\be
	\l_{hhh}\,\l_{h\varphi\varphi}=\l_{h\varphi\varphi}^2\,,
\ee
whose non-trivial solution is only\footnote{The other solution, $\l_1=0$ corresponds to a PM field that does not have any interactions with itself and Gravity, hence cannot interact with matter that couples to Gravity in a generally covariant manner.} 
\be
	\l_{h\varphi\varphi}=\l_{hhh}\,, 
	\label{l1l0}
\ee
which is consistent with the universality of gravitational interactions.

The resulting global symmetry algebra in $D>3$ is $\mathfrak{so}(1,D+1)$, which includes the $dS_D$ isometry subalgebra $\mathfrak{so}(1,D)$ generated by $M^{\sst AB}$, the Killing tensors of massless spin two field.
The additional generators $K^{\sst A}$ which uplift   $\mathfrak{so}(1,D)$ to $\mathfrak{so}(1,D+1)$
correspond to the Killing tensors of PM spin two field. 
It is interesting to note that when $D=3$ we get $\mathfrak{iso}(1,3)$ instead of $\mathfrak{so}(1,4)$
as the global symmetry algebra.\footnote{Interestingly, when considering four dimensional Einstein Gravity with a $dS_3$ slicing of the asymptotically Minkowski space with $\mathfrak{iso}(1,3)$ global symmetry, one encounters the ``massless'' and ``PM'' degrees of freedom on $dS_3$ slices \cite{Compere:2011ve,Troessaert:2017jcm}. This might indicate that the PM Gravity in three dimensions might be a consistent sector of Einstein-Hilbert Gravity with zero cosmological constant in four dimensions.}

In four dimensions, we recover the results of \cite{Joung:2014aba}: the corresponding algebra is $\mathfrak{so}(1,5)$.
Note that if we consider $AdS_4$ instead of $dS_4$, there are again two possible real forms: $\mathfrak{so}(2,4)$, corresponding to conformal gravity, and $\mathfrak{so}(3,3)$, which would correspond to a positive relative sign for kinetic terms of graviton and PM fields in $AdS_4$ background.\footnote{It was noted in \cite{Vasiliev:2007yc} that the free massless fields in $AdS_4$ can manifest not only conformal symmetry $\mathfrak{so}(2,4)\sim \mathfrak{su}(2,2)$ but also $\mathfrak{so}(3,3)\sim \mathfrak{sl}(4,\mathbb R)$. A question was raised there whether such a symmetry can be extended to a non-linear theory. That question would be equivalent to the one this work is attempting to answer --- whether there is a consistent non-linear theory with $\mathfrak{so}(1,5)\sim \mathfrak{su}^*(4)$ symmetry in $dS_4$. }
The generalisation of this statement to arbitrary $D>3$ is straightforward: in $AdS_D$, relative negative sign corresponds to familiar conformal gravity with a global symmetry $\mathfrak{so}(2,D)$, while relative positive sign of PM and massless spin two corresponds to the global symmetry algebra $\mathfrak{so}(3,D-1)$.
Perturbative unitarity will not be possible in that case though, since the PM field is not unitary in $AdS_D$.

\paragraph{Admissibility condition}

Finally, we can examine the admissibility condition. The isometry --- corresponding to \eqref{isometry} ---
and the $\mathfrak{so}(1,D)$ covariance of $K^{\sst A}$ --- corresponding to \eqref{general cov} ---
are simply inherited from their gauge versions, namely the diffeomorphism and general covariance of $\varphi_{\m\n}$.
They do not cause any problem of the admissibility condition, as we understand that it is straightforward to write an
action $S[h,\varphi]$ in such a manner. The potential problem is in the PM gauge transformation, 
which translates here to the question whether the bracket \eqref{PM com} is correctly represented 
in terms of $\delta^{\sst [1]}$. By computing the relevant commutators, we obtain
\ba
	&& \left(\delta^{\sst [1]}_{\bar \a_{2}}\,\delta^{\sst [1]}_{\bar \a_{1}}
	-\delta^{\sst [1]}_{\bar \a_{1}}\,\delta^{\sst[1]}_{\bar \a_{2}}\right)h_{\m\n}
	\sim \d^{\sst [1]}_{[\![ \bar\a_1,\bar\a_2]\!]}\,h_{\m\n}\,, \\
	\label{CommH}
	&& \left(\delta^{\sst [1]}_{\bar \a_{2}}\,\delta^{\sst [1]}_{\bar \a_{1}}
	-\delta^{\sst [1]}_{\bar \a_{1}}\,\delta^{\sst[1]}_{\bar \a_{2}}\right)\varphi_{\m\n}
	=
	\d^{\sst [1]}_{[\![ \bar\a_1,\bar\a_2]\!]}\,\varphi_{\m\n}
	+(\l_{h\vf\vf}^2+\l_{\varphi\varphi\varphi}^2)\,\mathcal C_{\m\n}\,,
	\label{CommPhi}
\ea
where $\mathcal C_{\m\n}$ is given by
\be
	\mathcal C_{\m\n}= 
	4\,\Big[\partial^\rho\alpha_{[1}\,\partial^\s\alpha_{2]}\,\nabla_{(\m|}\nabla_{\s}\varphi_{|\nu)\rho}
	+\frac{2\,\L}{D-2}\,\a_{[1}\partial^\r\a_{2]}(\nabla_{(\m}\varphi_{\n)\r}-\nabla_\r\varphi_{\m\n})\Big]\,.
	\label{cC}
\ee
Therefore, the admissibility condition requires 
\be
	\l_{h\vf\vf}^2+\l_{\varphi\varphi\varphi}^2=0\,.
	\label{hvfvf}
\ee
Remind that $\l_{\varphi\varphi\varphi}$ is non-vanishing only in $D=4$ dimensions since the associated  vertex $\cV_{\varphi\varphi\varphi}$
exists only there.
Anyway,
 the above condition does not admit 
any non-trivial real solution in any dimensions.
This is the summary of the no-go theorem reported in \cite{Joung:2014aba}.

\section{Relaxing the Assumption on the Field Content}
\label{sec: enlarge FC}

Let us recapitulate everything one more time with a few new remarks:
a generally covariant unitary theory, involving massless and a PM spin two fields in $D\geq 4$ dimensional de Sitter space,
should have the Lie algebra $\mathfrak{g}=\mathfrak{so}(1,D+1)$ as its global symmetry.
One important point here is that this conclusion is valid
even if the theory contains additional fields besides the massless and PM spin two fields.
If this additional part includes other gauge fields, then the actual global symmetry
will be enlarged in a way to include $\mathfrak{g}$ as a subalgebra.
Assuming that there is no other field than the massless and PM spin two fields,
we derived the transformation $\delta^{\sst [1]}_{\bar\alpha}$ acting on the on-shell fields $h$ and $\varphi$
and verified that it does not close, in other words, it is incapable of forming a representation of $\mathfrak{g}$.
At this last point, namely addressing the admissibility condition, it is natural to ask: what would change if we include other fields, say $\phi_i$?
For the moment, $\phi_i$ can be any kind of fields.

Let us revisit our general analysis of Section \ref{sec: gauge invariance}
and Section \ref{sec: PM no-go}
adding new fields $\phi_i$. Then, 
we should take into account the following new cubic interactions.
\ba
	 h\!-\!h\!-\!\phi_i\,, \qquad &h\!-\!\varphi\!-\!\phi_i\,, 
	\qquad &\varphi\!-\! \varphi\!-\! \phi_i\,,\nn
	h\!-\! \phi_i\!-\! \phi_j\,,\qquad &\varphi\!-\! \phi_i\!-\!\phi_j\,,
	\qquad
	&\phi_i\!-\!\phi_j\!-\!\phi_k\,.
\ea
For now, we consider only the cubic interactions which induce
non-trivial $\delta^{\sst [1]}$.
The third coupling of the second line does not contain any massless or PM field, hence
does not influence the consistency at least at this order.
The general covariance of the gauge algebra, or the $\mathfrak{so}(1,D)$ covariance
of the global algebra, 
would forbid the first two couplings in the first line
 whereas
allow only the diagonal one with $i=j$ for the first coupling in the second line.
This diagonal $h\!-\!\phi_i\!-\!\phi_i$ coupling is simply the gravitational minimal coupling of the field $\phi_i$.
The obstruction we want to resolve is in the $\bar\a$ transformation, and the 
relevant cubic interactions are the third one in the first line and the second one in the second line, 
which we rewrite here with coupling constants as
\be
	S^{\sst [3]}[\varphi,\phi_i]
	=\int d^4\sqrt{-\bar g}\,\Big(
	\sum_{i}\g_i\,\cW_i(\varphi,\varphi,\phi_i)
	+\sum_{i,j}\g_{ij}\,\cW_{ij}(\varphi,\phi_i,\phi_j)\Big)\,.
\ee
These couplings can induce new terms in the $\delta^{\sst [1]}_{\bar\a}$ transformation as
\be
	\g_i\,\cW_i(\varphi,\varphi,\phi_i)
	\quad\longrightarrow\quad 
	\left\{
	\begin{array}{c}
	\ \left[	\delta^{\sst [1]}_{\bar\a} \varphi\right]_{\phi_i} =\g_i\,R(\bar\a)^{\varphi}{}_{\phi_i}\,\phi_i
		\vspace{5pt}\\
	\ \left[	\delta^{\sst [1]}_{\bar\a} \phi_i\right]_{\varphi} =\g_i\,R(\bar\a)^{\phi_i}{}_{\varphi}\,\varphi
	\end{array}\right.\,,
	\label{2PM}
\ee
and 
\be
	\g_{ij}\,\cW_{ij}(\varphi,\phi_{i},\phi_j)
	\quad\longrightarrow\quad 
	\left\{
	\begin{array}{c}
	\ \left[	\delta^{\sst [1]}_{\bar\a} \phi_i\right]_{\phi_j} =\g_{ij}\,R(\bar\a)^{\phi_i}{}_{\phi_j}\,\phi_j
	\vspace{5pt}\\
	\ \left[	\delta^{\sst [1]}_{\bar\a} \phi_j\right]_{\phi_i} =\g_{ij}\,R(\bar\a)^{\phi_j}{}_{\phi_i}\,\phi_i
	\end{array}\right.\,,
	\label{1PM}
\ee
Note that the above formulas are schematic:
$R(\bar \a)^B{}_A$ are certain differential operators sending the tensor field $A$ to the tensor field $B$.
Their precise forms are dictated by the cubic vertices $\cW_i(\varphi,\phi_{i},\phi_j)$.
With these structures in mind, let us come back to our problem:
non-closure \eqref{CommPhi} of the $\delta^{\sst [1]}_{\bar\a}$ transformation on $\varphi$ 
(remind that the $\delta^{\sst [1]}_{\bar\a}$ closes on $h$: see \eqref{CommH}).
It is clear that among \eqref{2PM} and \eqref{1PM}, only
the former may have a chance to fix this problem, so let us focus on it:
the latter will become equally important in subsequent analysis.
The introduction of the $\varphi\!-\! \varphi\!-\! \phi_i$ cubic interactions will alter 
 the $\delta^{\sst [1]}_{\bar\a}$ commutator by
\be
	\left[\left(\delta^{\sst [1]}_{\bar \a_{1}}\,\delta^{\sst [1]}_{\bar \a_{2}}
	-\delta^{\sst [1]}_{\bar \a_{2}}\,\delta^{\sst[1]}_{\bar \a_{1}}\right)\varphi_{\m\n}\right]_{\phi_i}
	=
	 \g_i^2\left(
	\big[R(\bar\a_2)^{\varphi}{}_{\phi_i}\,R(\bar\a_1)^{\phi_i}{}_{\varphi}-
	R(\bar\a_1)^{\varphi}{}_{\phi_i}\,R(\bar\a_2)^{\phi_i}{}_{\varphi}\big]\,\varphi\right)_{\m\n}\,,\label{com change}
\ee
where  the notation $[\,\cdots]_{\phi_i}$ indicates the terms arising through the field $\phi_i$.
The point is whether the right hand side can compensate the problematic $\cC_{\m\n}$ terms.
In other words whether
$R(\bar\a_{[1})^{\varphi}{}_{\phi_i}\,R(\bar\a_{2]})^{\phi_i}{}_{\varphi}$ can make 
precisely the same structure as $\cC_{\m\n}$ in \eqref{cC}.
If it is the case, then we may solve the problem of the equation \eqref{hvfvf} 
as it will be modified by a term proportional to $\g_i^2$.

A priori, the field $\phi_i$ can be of any type, but as we will argue below, only symmetric rank two tensor will 
be capable of reproducing the structure like $\cC_{\m\n}$.
To understand this, let us first assume $\phi_i$ be a rank $r$ tensor ($\phi_i=\phi_{\rho_1\cdots\rho_r}$) of any symmetry type,
and write an ansatz for $\left[\delta^{\sst [1]}_{\bar\a} \varphi\right]_{\phi_i}$ and 
$\left[\delta^{\sst [1]}_{\bar\a} \phi_i\right]_{\varphi}$
 as
\ba
	&& \Big(R(\bar\a)^{\varphi}{}_{\phi_i}\,\phi_i\Big)_{\m\n}
	=A^{\rho_1\cdots \rho_r}_{\m\n}(\bar g,\nabla_1,\nabla_2)\,\bar\alpha(x_1)\,\phi_{\rho_1\cdots \rho_r}(x_2)\,\Big|_{x_1=x_2}\,,\nn
	&& \Big(R(\bar\a)^{\phi_i}{}_{\varphi}\,\varphi\Big)_{\m_1\cdots \m_r}
	= B_{\m_1\cdots \m_r}^{\rho_1\rho_2}(\bar g,\nabla_1,\nabla_2)\,\bar\alpha(x_1)\,\varphi_{\rho_1\rho_2}(x_2)\,\Big|_{x_1=x_2}\,,
\ea
Here, $A^{\rho_1\cdots \rho_r}_{\m\n}$ and $B_{\m_1\cdots \m_r}^{\r_1\r_2}$ are tensors made out of the metric $\bar g_{\m\n}$
and the derivatives $\nabla_{1,\m}$ and $\nabla_{2,\m}$ acting respectively on 
the parameter $\bar\alpha(x_1)$ and the field $\phi_{\rho_1\cdots \rho_r}(x_2)$ or $\varphi_{\r_1\r_2}(x_2)$.
Let us check how the indices of $A^{\rho_1\cdots \rho_r}_{\m\n}, B_{\m_1\cdots \m_r}^{\r_1\r_2}$ can be distributed to  
$\bar g_{\m\n}$, $\nabla_{1,\m}$
and $\nabla_{2,\m}$.
Restrictions are imposed by the on-shell conditions of the field and the Killing conditions of the parameter:
$\nabla^{\rho_m}_2$ and $\bar g^{\rho_m\rho_n}$ are forbidden by the traceless-transverse (TT) conditions of the field,
the double derivative
$\nabla^{\rho_m}_1\,\nabla^{\rho_n}_1$ is forbidden because the Killing condition would replace it with $\bar g^{\rho_m\rho_n}$
which is forbidden. 
With these,
it is straightforward to see that for $r=0$ and $r\ge 4$, there are no candidate structures.
There are three remaining possibilities $r=1,2,3$.
Here, let us consider another condition:
since $\cC_{\m\n}$ \eqref{cC} has a four derivative term,
the product of  $A^{\rho_1\cdots \rho_r}_{\m\n}$
and $B_{\m_1\cdots \m_r}^{\r_1\r_2}$  should have a four derivative term as well.
For $r=1$, such a term can arise only from a single possibility, 
\be
		A^{\rho}_{\m\n} \sim \delta^{\rho}_{(\mu}\,\nabla_{2,\n)}\,\nabla_{1}^{\lambda}\,\nabla_{2,\lambda}\,,\qquad
	B^{\r_1\r_2}_\m \sim \delta^{(\rho_1}_{\mu}\,\nabla_1^{\rho_2)}\,,
	\label{spin 1}
\ee
For $r=3$, it is also unique up to a symmetrization of the indices of the rank 3 tensor
(that is $\rho_1\rho_2\rho_3$ and $\m_1\m_2\m_3$):
\be
	A^{\rho_1\rho_2\rho_3}_{\m\n}\sim \delta^{\rho_1}_{(\mu}\delta^{\rho_2}_{\nu)}\,\nabla_1^{\rho_3}\,,
	\qquad
	B^{\rho_1\rho_2}_{\m_1\m_2\m_3} \sim \delta^{(\rho_1}_{\mu_1}\delta^{\rho_2)}_{\m_2}\,
	\nabla_{2,\m_3}\,\nabla_{1}^{\lambda}\,\nabla_{2,\lambda}\,.
	\label{spin 3}
\ee
Note that the derivatives in \eqref{spin 1} and \eqref{spin 3} are distributed unevenly in the tensors $A$ and $B$
even though both should be associated with a single cubic interaction. This is not a contradiction and they may indeed arise from a three- or five- derivative coupling.
However, such couplings to odd-spin fields  
vanish identically up to total derivative term by the symmetry of two PM fields involved in the cubic vertex.
Finally for $r=2$, there are two terms which may cancel the four derivative term in $\cC_{\m\n}$\,:
up to a relative factor and a symmetrization of $\r_1\r_2$ and $\m_1\m_2$, they are
\be
	A^{\rho_1\rho_2}_{\m_1\m_2},\ B^{\rho_1\rho_2}_{\m_1\m_2}
	\ \sim\ \delta^{\rho_1}_{\mu_1}\,\nabla_1^{\rho_2}\,\nabla_{2,\m_2}
	+\delta^{\rho_1}_{\m_1}\delta^{\rho_2}_{\m_2}\,\nabla_{1}^{\lambda}\,\nabla_{2,\lambda}
	\,.
	\label{A B 2}
\ee
Let us examine this possibility in more detail:
\be
	\cC_{\m_1\m_2} \propto A^{\rho_1\rho_2}_{\m_1\m_2}(\nabla_{1},\nabla_{2}+\nabla_{3})\,\bar\a_{[1}(x_1)\,
	B_{\r_1\r_2}^{\n_1\n_2}(\nabla_2,\nabla_3)\,\bar\a_{2]}(x_2)\,\varphi_{\n_1\n_2}(x_3)\,\Big|_{x_1=x_2=x_3}\,.
	\label{C=AB}
\ee
It will be  convenient to consider the contraction 
of the above equation with an arbitrary symmetric tensor $\tilde \varphi_{\m\n}$. 
Using the precise form \eqref{cC} of $\cC_{\m_1\m_2}$ and discarding a total derivative term, we find
the four-derivative part of $\tilde \varphi^{\m\n}\,\cC_{\m\n}$ is proportional to
\be
	\left(\partial^\m \bar \a_{[1}\,\nabla_{\m}\tilde \varphi^{\r_1\r_2}\right) 
	\left(\partial^\n \bar \a_{2]}\,\nabla_{(\r_1}\,\varphi_{\r_2)\n}\right),
	\label{CC}
\ee
whereas the contraction of the right hand side of \eqref{C=AB} with $\tilde \varphi_{\m\n}$ reads
\be
	\left[A^{\rho_1\rho_2}_{\m_1\m_2}(\nabla_{1},\nabla_{2})\,\bar\alpha_{[1}\,\tilde\varphi^{\m_1\m_2}\right]
	\left[B^{\n_1\n_2}_{\r_1\r_2}(\nabla_{1},\nabla_{2})\,\bar\alpha_{2]}\,\varphi_{\n_1\n_2}\right].
	\label{AB}
\ee
Inside the square brackets,
the identification $x_1=x_2$ is understood.
By requiring the proportionality between \eqref{CC} and \eqref{AB}, 
we find that the choice
\be
	A^{\rho_1\rho_2}_{\m_1\m_2}\sim
	\delta^{\rho_1}_{\m_1}\delta^{\rho_2}_{\m_2}\,\nabla_{1}^{\lambda}\,\nabla_{2,\lambda}\,,
	\qquad
	B^{\n_1\n_2}_{\r_1\r_2}
	\ \sim\ \delta^{\n_1}_{\r_1}\,\nabla_1^{\n_2}\,\nabla_{2,\r_2}\,.
	\label{A B sol}
\ee
gives the structure \eqref{CC}.
From the structure of $A^{\rho_1\rho_2}_{\m_1\m_2}$,
we also find the intermediate field $\phi_{\rho_1\rho_2}$
is a symmetric tensor.
Therefore, only symmetric rank two, namely  spin two fields
have a chance to cure the problem of $\delta^{\sst [1]}_{\bar\a}$ non-closure on the PM field. 

\subsection{Adding Spin-Two Fields}
\label{sec: spin2 tower}

Let us study the case we add one spin two field $\phi_{\m\n}$. 
As we discussed in \eqref{admin cond}, we shall also impose the on-shell condition \eqref{OS},
but the mass value of $\phi_{\m\n}$ is not determined yet.
Interestingly, the on-shell condition alone restricts severely possible form of $\left[\delta^{\sst [1]}_{\bar\a} \varphi_{\m\n}\right]_{\phi}$
and
$\left[\delta^{\sst [1]}_{\bar\a} \phi_{\m\n}\right]_{\varphi}$.
In order to appreciate this point, let us do the analysis for two fields $\chi_{\m\n}$ and $\psi_{\m\n}$
with arbitrary mass values.
Eventually, we will apply the result of the analysis 
to the problem of adding a new field $\phi_{\m\n}$ to the system of massless and PM field.
We can first write the most general form of the relevant $\delta^{\sst [1]}_{\bar\a}$ as
\ba
	&&\left[\delta^{\sst [1]}_{\bar\a} \chi_{\m\n}\right]_{\psi} =
	a_1\,\frac12\,\nabla^\rho\,\bar\alpha\,\nabla_\rho\,\psi_{\m\n}
	+a_2\,\nabla^\rho\,\bar\alpha\,\nabla_{(\m}\,\psi_{\n)\rho}
	+a_3\,\frac{1}{2\,L^2}\,\bar\a\,\psi_{\m\n}\,,
	\label{pm to m} \\
	&& \left[	\delta^{\sst [1]}_{\bar\a} \psi_{\m\n}\right]_{\chi} =
	b_1\,\frac12\,\nabla^\rho\,\bar\alpha\,\nabla_\rho\,\chi_{\m\n}
	+b_2\,\nabla^\rho\,\bar\alpha\,\nabla_{(\m}\,\chi_{\n)\rho}
	+b_3\,\frac{1}{2\,L^2}\,\bar\a\,\chi_{\m\n}\,.
	\label{m to pm}
\ea
The first two terms of the right hand sides of \eqref{pm to m} and \eqref{m to pm}
are nothing but the structures in \eqref{A B 2}.
Here, we have also included the no-derivative last terms for a complete analysis.

\subsubsection{On-shell constraints}

There are three constraints on on-shell fields given through Fierz equations \eqref{OS}:
mass-shell, transversality and traceless conditions. The trace condition does not induce more conditions on the form of the global transformations \eqref{pm to m} and \eqref{m to pm} (we already chose these transformations to not involve traces of the fields involved) once we impose the transversality and Klein-Gordon equations.
 
The transversality condition on the fields imposes the relations,
\be
	D\,a_1+\left(D+2+s_\psi\right) a_2+a_3=0\,,\qquad 
	D\,b_1+\left(D+2+s_\chi\right)b_2+b_3=0\,,
	\label{trans}
\ee
where $s_\chi$ and $s_\psi$ are related to the mass values $m_\chi$ and $m_\psi$ by
\be
	s=m^2\,L^2\,.
	\label{mu values}
\ee
With this choice, the massless and PM spin two fields have $s=0$ and $D-2$, respectively. In the case of our main interest, we have $s_\chi=D-2$, but for the moment we can keep it arbitrary.
The tracelessness of the fields are compatible with \eqref{pm to m} and \eqref{m to pm}.
It remains to check the last on-shell condition:
\be
	\left(\Box-\frac{2+s_\chi}{L^2}\right) \left[	\delta^{\sst [1]}_{\bar\a} \chi_{\m\n}\right]_{\psi} = 0\,,
	\qquad
	\left(\Box-\frac{2+s_\psi}{L^2}\right) \left[	\delta^{\sst [1]}_{\bar\a} \psi_{\m\n}\right]_{\chi} = 0\,.	
\ee
The first of these conditions translate into
\begin{align}
2\,(2+s_\psi)\,a_1 + 4\,D\, a_2 +(D-s_\psi+s_\chi)\,a_3=0\,,
\label{os eq1} \\
(D-2+s_\psi-s_\chi)\,a_1+4\,a_2+2\,a_3=0\,,
\label{os eq2} \\
2\, a_1+(D+s_\psi-s_\chi)\,a_2=0\,.
\label{os eq3}
\end{align}
Together with \eqref{trans} we get four equations for three variables $a_1,a_2,a_3$. In order  to have non-trivial solutions, 
all the $3\times 3$ minors of the following matrix should be zero:
\be
\left(
\begin{array}{ccc}
 D & D+2+s_\psi & 1  \\
 2\,(2+s_\psi) & 4\,D & D-s_\psi+s_\chi  \\
 D-2+s_\psi-s_\chi & 4 & 2  \\
 2 & D+s_\psi-s_\chi  & 0  \\
\end{array}
\right)
\ee
Such condition is equivalent to
\be
	(s_\chi-s_\psi)^2+2\,(s_\chi+s_\psi)=D(D-2)\,.
	\label{s eq}
\ee
Note that the above equation is symmetric in the exchange of $s_\chi$ and $s_\psi$. 
This means that by examining the on-shell conditions for $b_1, b_2, b_3$, we obtain
exactly the same condition for $s_\chi$ and $s_\psi$.
For the solution of \eqref{s eq}, it is useful to parametrize $s$ as
\be
	s=\m\,(D-1-\m)\,,
	\label{mass para}
\ee
which is invariant under
\be
	\m \to D-1-\m\,.
	\label{flip}
\ee
The massless and PM spin two fields corresponds to the points $\m=0$ and $1$ 
(or $D-1$ and $D-2$).
The range $0<\mu<1$ (or $D-2<\mu<D-1$) is forbidden by unitarity,
while the range $1<\mu<D-2$ corresponds to a massive field, up to \eqref{flip}.
With the parametrization \eqref{mass para}, the condition \eqref{s eq} becomes 
\be
	(\m_\chi+\m_\psi-D)\,(\m_\chi+\m_\psi-D+2)\left(\m_\chi-\m_\psi-1\right)
	\left(\m_\chi-\m_\psi+1\right)=0\,.
\ee
For a given $\m_\chi$, the above clearly has  four solutions for $\m_\psi$:
\be
	\m_\psi=
	\left\{
	\begin{array}{c}
	D-2-\m_\chi\\
	D-\m_\chi\\
	\m_\chi+1\\
	\m_\chi-1\end{array}\right.,
	\label{os cond}
\ee
but the first two solutions are in fact related to the last two through \eqref{flip},
so it is sufficient to consider  the latter cases only.
From now on, let $\psi^\pm_{\m\n}$ denote the fields
with $\m_{\psi^\pm}=\m_\chi\pm 1$ for convenience.
Note however that this convention depends on the choice of $\mu_\chi$
from $s_\chi$:
if $\mu_\chi\to D-1-\m_\chi$, then $\psi^\pm_{\m\n} \to \psi^{\mp}_{\m\n}$.
In terms of $s$, 
the solution can be expressed as 
\be
	s_\chi=\m_\chi(D-1-\m_\chi)\,,\qquad
	s_{\psi^\pm}=(\m_\chi\pm1)(D-1\mp1-\m_\chi)\,.
\ee
Remark that the $\bar\alpha$-transformations \eqref{pm to m} and \eqref{m to pm}
are consistent only for the fields whose $\mu$ values differ by 1
up to \eqref{flip}. This is a manifestation of the fact that the PM field can interact with two spin-two fields through a cubic vertex, only if the mass values of the latter fields differ by one in terms of $\m$. In particular, the interaction with two identical fields is only possible for a specific mass, that can be written in two forms \eqref{mass para} with two values of $\m$ that differ by one. This mass value coincides with the PM mass in four dimensions, allowing for cubic self-interaction of PM field with two derivatives. 
We show in the Appendix \ref{sec: scalar coupling} that this pattern persists also for matter fields, providing a field-theoretical explanation of why Conformal Gravity chooses specific mass values for fields that can couple to it.

For the above values of $s_{\psi^\pm}$, or equivalently for $\m_{\psi^\pm}$ up to \eqref{flip},
the equations \eqref{trans}, \eqref{os eq1}, \eqref{os eq2}, \eqref{os eq3} and their $b_n$ counterparts
have a non-trivial solution,
\be
 \left(\begin{array}{c} a_1\\ a_2 \\ a_3 \end{array}\right)
 =a_+\, \left(\begin{array}{c} -(D-1-\m_\chi) \\ 1 \\ (D-\m_\chi)(D-3-\m_\chi) \end{array}\right),
 \qquad
 \left(\begin{array}{c} b_1\\ b_2 \\ b_3 \end{array}\right)= 
 b_+\, \left(\begin{array}{c} -(\m_\chi+1) \\ 1 \\ (\m_\chi-1)(\m_\chi+2) \end{array}\right),
 \label{+ os}
\ee
and 
\be
 \left(\begin{array}{c} a_1\\ a_2 \\ a_3 \end{array}\right)=
 a_-\, \left(\begin{array}{c} -\m_\chi \\ 1 \\ (\m_\chi-2)(\m_\chi+1) \end{array}\right),
 \qquad
 \left(\begin{array}{c} b_1\\ b_2 \\ b_3 \end{array}\right)
 =
 b_-\, \left(\begin{array}{c} -(D-\m_\chi) \\ 1 \\ (D+1-\m_\chi)(D-2-\m_\chi) \end{array}\right),
 \label{- os}
\ee
where $a_\pm$ and $b_\pm$ are undetermined constants.
It is worth to note that the $+$ coefficients and $-$ coefficients
are related by $\m_\chi\to D-1-\m_\chi$.
Let us also remark that 
the ratio $a_\pm/b_\pm$ is not physical since one can change it
by redefining the field $\chi_{\m\n}$ or $\psi^\pm_{\m\n}$ with a multiplicative factor. In the Lagrangian with fixed normalisation of these fields, both $a_\pm$ and $b_\pm$ are given through the same cubic coupling constant.
Hence, only the product $a_\pm\,b_\pm$ is remained to be determined.
As we discussed in the previous section around \eqref{2PM} and \eqref{1PM},
it is proportional to the square of the coupling constant of
the $\varphi\!-\!\chi\!-\!\psi^\pm$ cubic interaction. 
Note that the formula \eqref{+ os} does not reproduce \eqref{h varphi transf} for $\mu_\chi=0$. This is not an inconsistency though, as the difference lies in the lower order gauge transformation $\d^{\sst [0]}_\xi\, h_{\m\n}$ for the massless spin two field with a parameter $\xi_\m \propto \partial^\r\a \,\vf_{\m\r}$.

\subsubsection{Commutator}

Now let us examine the commutator of 
the transformations \eqref{pm to m} and \eqref{m to pm}.
Straightforward computations give
\ba
	&& \left[\delta^{\sst [1]}_{\bar \a_{2}}\,\delta^{\sst [1]}_{\bar \a_{1}}
	\,\chi_{\m\n}-\delta^{\sst [1]}_{\bar \a_{1}}\,\delta^{\sst [1]}_{\bar \a_{2}}
	\,\chi_{\m\n}\right]_{\psi}
	= c_1\,\frac12\,
	\partial^\rho\alpha_{[1}\,\partial^\s\alpha_{2]}\,\nabla_{(\m|}\nabla_{\s}\chi_{|\nu)\rho} +
		\label{commutes}\\
		&&\quad+\,\frac{1}{2\,L^2}
		\left(c_2\,
	\a_{[1}\partial^\r\a_{2]}\,\nabla_{(\m}\chi_{\n)\r}+c_3\,\frac12\,
	\a_{[1}\partial^\r\a_{2]}\,\nabla_\r\chi_{\m\n}+c_4\,
	\partial^\rho\alpha_{[1}\,\partial_{(\m}\alpha_{2]}\,\chi_{\n)\r}\right)
	\nonumber
\ea
with
\be
	C=\left(\begin{array}{c} c_1 \\ c_2 \\ c_3 \\ c_4 \end{array}\right)
	=\left(\begin{array}{c} 
	 -a_1\,b_2+a_2\,b_1-a_2\,b_2 \\
	  a_1\,b_2+a_2\,b_1+a_2\,b_2-a_2\,b_3+a_3\,b_2 \\
	a_1\,b_1+2\,a_2\,b_2+a_3\,b_1-a_1\,b_3 \\
	-a_1\,b_1-2\,a_1\,b_2+a_2\,b_3 \end{array}\right).\label{C}
\ee
As we showed in the previous section,
the on-shell conditions restrict $\psi_{\m\n}$
to two possibilities: $\psi^\pm_{\m\n}$ with  $\m_{\psi^\pm}=\m_\chi\pm1$\,.
About $\psi^+_{\m\n}$,  
the $a_n$ and $b_n$ coefficients are constrained as \eqref{+ os}
and the commutator structure is given with $C=C_+$ where
\be
	C_+=a_+\,b_+ \left(D-3-2\,\mu_\chi\right)
	\left(\begin{array}{c} 1 \\ 
	D-1 \\ 
	\m_\chi^2-(D-2)\,\m_\chi-D-1 \\ 1-\mu_\chi \end{array}\right).
	\label{C+}
\ee
About $\psi^-_{\m\n}$,
the coefficients in \eqref{- os} gives $C=C_-$ where
\be
	C_-=-\,a_-\,b_- \left(D+1-2\,\mu_\chi\right)
	\left(\begin{array}{c} 1 \\ 
	D-1 \\ \m_\chi^2-D\,\m_\chi-2 \\ \m_\chi-D+2\end{array}\right).
	\label{C-}
\ee
Finally, the entire commutator on the field $\chi_{\m\n}$ is simply
the sum of the $\psi^+_{\m\n}$ and the $\psi^{-}_{\m\n}$ contributions:
\be
	\delta^{\sst [1]}_{\bar \a_{[2}}\,\delta^{\sst [1]}_{\bar \a_{1]}}
	\,\chi_{\m\n}=
	\left[\delta^{\sst [1]}_{\bar \a_{[2}}\,\delta^{\sst [1]}_{\bar \a_{1]}}
	\,\chi_{\m\n}\right]_{\psi^+}
	+
	\left[\delta^{\sst [1]}_{\bar \a_{[2}}\,\delta^{\sst [1]}_{\bar \a_{1]}}
	\,\chi_{\m\n}\right]_{\psi^-}\,,
\ee
and its expression is given as the right hand side of \eqref{commutes}
with $C=C_++C_-$\,.

\subsection{Spin-two tower}

Let us apply the result obtained in the previous section to
the system of massless and PM spin two fields.
If we take $\chi_{\m\n}$ as the massless field $h_{\m\n}$, then $\psi^+_{\m\n}$
is the PM field $\varphi_{\m\n}$ whereas $\psi^-_{\m\n}$ becomes tachyonic. 
Therefore, we disregard the link to $\psi^-_{\m\n}$ by setting $a^h_-=0$
(here, the field under consideration is indicated by the superscript).
Then $C^h_-=0$ and
\be
	C^h=C^h_+
	=a^h_+\,b^h_+ \left(D-3\right)
	\left(\begin{array}{c} 1 \\ 
	D-1 \\ -D-1 \\ 1\end{array}\right).
\ee
For the closure of the symmetry, the commutator
should reproduce a $\mathfrak{so}(1,D)$ isometry transformation,
that is
the Lie derivative with the Killing tensor 
\be
	\bar \xi_\m=[\![\bar\a_2,\bar\a_1]\!]_\m=-\frac{2(D-3)}{L^2}\,\l_{h\varphi\varphi}\,\bar\a_{[2}\,\partial_\m\,\bar\a_{1]}\,.
	\label{ComKillA}
\ee
This requirement translates to the vector $C$ as $C=C_{\star}$ where
\be
	C_\star=2\,(D-3)\,\l_{hhh}\,\l_{h\varphi\varphi}
	\left(\begin{array}{c} 0 \\ 
	0 \\ -1 \\  1\end{array}\right).
	\label{C star}
\ee
At first look,  $h_{\m\n}$ does not seem to satisfy this requirement since $C^h$ differs from $C_\star$.
In fact, for the massless field $h_{\m\n}$, one should also take into account the linearised gauge transformation
$\delta^{\sst [0]}_{\varepsilon}$ with a field dependent gauge parameter $\varepsilon_\m$.
Indeed, the following full gradient terms
can contribute to the commutator:
\ba
	&& \nabla_{(\m}\left(\partial^\r\,\a_{[1}\,\partial^\s\,\a_{2]}\,\nabla_\s\,h_{\n)\rho}\right)=\nn
	&&
	\quad=\,\partial^\rho\alpha_{[1}\,\partial^\s\alpha_{2]}\,\nabla_{(\m|}\nabla_{\s}h_{|\nu)\rho} +
	\frac{1}{L^2}
	\,\a_{[1}\,\partial^\r\,\a_{2]}\left(\nabla_{(\m}\,h_{\n)\rho}-\nabla_\rho\,h_{\m\n}\right),\nn
	&&
	\nabla_{(\m}\left(\a_{[1}\,\partial^\r\,\a_{2]}\,h_{\n)\rho}\right)
	=\a_{[1}\,\partial^\r\,\a_{2]}\,\nabla_{(\m}\,h_{\n)\rho}
	-\partial^\r\,\a_{[1}\,\partial_{(\m}\,\a_{2]}\,h_{\n)\rho}\,,
\ea
and we should quotient them out.
In terms of the vector $C$, this amounts to imposing
the equivalence relation,
\be
	C^h\sim C^h+
	c\,\left(\begin{array}{c} 1 \\ 
	1 \\ -2 \\  0\end{array}\right)
	+d\,\left(\begin{array}{c} 0 \\ 
	1 \\ 0 \\  -1\end{array}\right),
\ee
where $c$ and $d$ are arbitrary constants.
With the above, we can achieve $C^h\sim C_\star$ upon 
the identification, 
\be
	a_+^h\,b_+^h=\frac{2\,\l_{hhh}\,\l_{h\varphi\varphi}}{D-1}\,.
	\label{ah bh}
\ee
Therefore, the transformation rule between $h_{\m\n}$ and $\varphi_{\m\n}$
is completely fixed by asking the on-shell condition and the closure of the 
global symmetry. 
The result coincides with the one \eqref{h varphi transf}
obtained from the gravitational minimal coupling of $\varphi_{\m\n}$.

Now we move to the next player, the PM field $\varphi_{\m\n}$.
Taking the $\chi_{\m\n}$ field as $\varphi_{\m\n}$,
the other two fields
$\psi^\pm_{\m\n}$ becomes the massless field $h_{\m\n}$ and
 a new massive  field $\phi_{\m\n}$, respectively.
By asking $C^\varphi=C_\star$, we get
\be
	(D-5)\,a_+^\varphi\,b_+^\varphi=(D-1)\,a_-^\varphi\,b_-^\varphi=2\,\l_{h\varphi\varphi}^2\,.
\ee
Note that $a_-^\varphi=b_+^h$ and $b_-^\varphi=a_+^h$ and 
the above is consistent with \eqref{ah bh} since 
$\l_{hhh}=\l_{h\varphi\varphi}$ \eqref{l1l0}.

Since we added the new field $\phi_{\m\n}$, we have to 
require
the closure of the symmetry on the new field as well.
Similarly to the $h_{\m\n}$ and $\varphi_{\m\n}$ cases, 
the consistency on $\phi_{\m\n}$ would require to introduce
yet another spin-two field.
In this way, we involve a multitude of spin-two fields recursively.
For a clear organization of this tower of fields, we 
label all the fields involved as $\phi^{\sst (n)}_{\m\n}$\,,
where  $h_{\m\n}=\phi^{\sst (0)}_{\m\n}$\,, $\varphi_{\m\n}=\phi^{\sst (1)}_{\m\n}$
and the newly added field $\phi_{\m\n}=\phi^{\sst (2)}_{\m\n}$\,.
In this notation, the mass value of $\phi^{\sst (n)}_{\m\n}$ is given with
$\m_{\phi^{\sst (n)}}=n$.
By taking now $\chi_{\m\n}$ as $\phi^{\sst (n)}_{\m\n}$
and $\psi^\pm_{\m\n}$ as $\phi^{\sst (n\pm1)}_{\m\n}$,
the requirement $C^{\sst (n)}=C_\star$ gives 
\be
	a^{\sst (n)}_\pm\,b^{\sst (n)}_\pm=\frac{2\,(D-3)\,\l_{hhh}^2}{(D-1-2\,n)(D-1-2(n\pm1))}\,,
	\label{ab gen}
\ee
which is consistent with
$a^{\sst(n)}_\pm=b^{\sst (n\pm1)}_\mp$.
In this way, we can include
more and more fields $\phi^{\sst (n)}_{\m\n}$ starting from $n=0$
and their $\bar\a$-transformation rule is determined by \eqref{+ os} and \eqref{- os} with \eqref{ab gen}.
However, when $n$ becomes sufficiently large, we encounter a subtlety.

\paragraph{In odd dimension $\bm D$,} when $D>3$ and $n$ reaches the value $(D-3)/2$ the coefficient $a^{\sst (n)}_+\,b^{\sst (n)}_+$ in \eqref{ab gen} diverges.
This is due to the fact that the commutator between $\varphi^{(\frac{D-3}2)}_{\m\n}$
and $\varphi^{(\frac{D-1}2)}_{\m\n}$ vanishes: $C_+^{(\frac{D-3}2)}=0$ in \eqref{C+}.
Note that the field $\varphi^{(\frac{D-1}2)}_{\m\n}$ is the special one with the symmetry $\m^{(\frac{D-1}2)}=D-1-\m^{(\frac{D-1}2)}$,
or in other words, it is the heaviest field among the ones with a real $\mu$.
The commutator between $\varphi^{(\frac{D-3}2)}_{\m\n}$ and $\varphi^{(\frac{D-5}2)}_{\m\n}$
alone cannot satisfy the requirement of the symmetry closure because $C^{(\frac{D-3}2)}_-\neq C_\star$\,, see \eqref{C-} and \eqref{C star}.
Therefore, in this case, the admissibility condition cannot be rescued. 
We shall comment on a potential resolution to this problem together with the $D=3$ case in the next section.

\paragraph{In even dimension $\bm D$,} when $n$ reaches the value $(D-2)/2$, 
the field $\varphi^{(\frac{D-2}2)}_{\m\n}$ will be linked to $\varphi^{(\frac{D}2)}_{\m\n}$
for the closure of the symmetry. Here, it is important to note that these two fields
have in fact the same mass value: $s^{(\frac{D-2}2)}=s^{(\frac{D}2)}=\frac{D}2(\frac{D}2-1)$. 
Moreover, the transformation from  $\varphi^{(\frac{D-2}2)}_{\m\n}$ to $\varphi^{(\frac{D}2)}_{\m\n}$
given by the coefficients $a_n$
and the transformation from $\varphi^{(\frac{D}2)}_{\m\n}$ to $\varphi^{(\frac{D-2}2)}_{\m\n}$
given by the coefficients $b_n$ are the same
up to an overall non-physical factor:
\be
 \left(\begin{array}{c} a_1\\ a_2 \\ a_3 \end{array}\right)
 =a_+\, \left(\begin{array}{c} -\frac D2 \\ 1 \\ (\frac D2+1)(\frac D2-2) \end{array}\right),
 \qquad
 \left(\begin{array}{c} b_1\\ b_2 \\ b_3 \end{array}\right)= 
 b_+\, \left(\begin{array}{c} -\frac D2 \\ 1 \\ (\frac D2+1)(\frac D2-2) \end{array}\right).
\ee
Given this, we will consider two possibilities,
either to identify the two fields or to proceed with the doubled spectra. 
We will discuss these two possibilities in the next section. 

These two possibilities are natural to consider, even though they are not the only possibilities. In fact, one can introduce an arbitrary number of fields with the mass value corresponding to $s=\frac{D}{2}(\frac{D}{2}-1)$. Such a construction may eventually lead to a theory with a spectrum equivalent to multiple copies of Conformal Gravity. As we will see, in any of these cases at least one of the fields has to be a ghost. We will restrict ourselves to the study of two possibilities mentioned above as the other options will not result in conceptually different models.

\subsection{Consistent Solutions}
\subsubsection{Conformal Gravity}
\label{Conformal Gravity}

In even dimensions, we can identify the fields
$\phi^{(\frac{D-2}2)}_{\m\n}$ and $\phi^{(\frac{D}2)}_{\m\n}$\,.
Then, the recursive introduction of new fields ends with $\phi^{(\frac{D-2}2)}_{\m\n}$,
which is mapped to itself under $\bar\alpha$-transformation: 
the coefficients $a^{(\frac{D-2}2)}_+$ and $b^{(\frac{D-2}2)}_+$ is also unified as  $a^{(\frac{D-2}2)}_+=b^{(\frac{D-2}2)}_+$\,. 
This leads to the condition,
\be
	\Big(a^{(\frac{D-2}2)}_+\Big)^2=-2\,(D-3)\,\l_{hhh}^2\,.
	\label{a eq}
\ee
Since the right hand side is negative, the above equation 
does not have any real solution $a_+^{(\frac{D-2}2)}$\,.
However, if we consider the case where $\phi^{\sst (n)}$ 
have alternating kinetic term signs, then 
the right hand side of \eqref{a eq} will acquire an additional minus sign
and the above equation admits  a real solution.
In such a case  the global symmetry  becomes
the conformal symmetry $\mathfrak{so}(2,D)$.
The resulting theory is nothing but the conformal gravity in $D$ dimensions.
It is usually written as a $D$-derivative theory of a single metric field, but
can be decomposed into a theory of Einstein gravity coupled to 
spin-two {\it matter} fields of $\m=1,\ldots,\frac{D-2}2$\,. Here, the only gauge fields are the massless and PM fields 
$\phi^{\sst (0)}_{\m\n}$ and $\phi^{\sst (1)}_{\m\n}$\,, and  the cubic interactions associated to the global symmetry transformations are
the ones involving at least one gauge field among the three fields entering the interaction.
These interactions are:
\be
	\phi^{\sst (0)}\!-\!\phi^{\sst (n)}\!-\!\phi^{\sst (n)}\,,\qquad
	\phi^{\sst (1)}\!-\!\phi^{\sst (n)}\!-\!\phi^{\sst (n+1)}\,,\qquad
	\phi^{\sst (1)}\!-\!\phi^{(\frac{D-2}2)}\!-\!\phi^{(\frac{D-2}2)}\,.
\ee
where $n=0,\ldots, \frac{D-2}2$ and some interactions appear more than once in this list.
When $D=4$, the last cubic interaction is the $\phi^{\sst (1)}_{\m\n}=\varphi_{\m\n}$ self-interaction
which exists only in four dimensions. 

In odd dimensions, so far there is no obvious candidate for conformal gravity except for the $D=3$ case
where we have the Chern-Simons realization with $\mathfrak{so}(2,3)$ gauge algebra \cite{Horne:1988jf}.
Recently the global conformal invariants --- the  scalar densities of  metric field  invariant under local Weyl rescalings up to a total derivative --- are classified both in even and odd dimensions \cite{Boulanger:2018rxo}. This classification has a good chance to provide decent candidates for what we can call as conformal gravity in dimensions $D=4\,n-1$. 
These odd dimensional candidates for conformal gravity have a parity odd structure, as is obvious in $3D$ Chern-Simons case.
In fact, the 3D conformal gravity can also be decomposed into the massless and PM fields, but 
it has a noteworthy oddity compared to the even dimensional theories:\footnote{Another
special feature is that the PM spin two has no-bulk DoF and only two boundary DoF,
which cannot be realized by a  two-derivative action but by a one-derivative one.
In fact, in 3D all maximal depth partially-massless fields of any spins admit two descriptions:
one is the extrapolation of $D>3$ case to $D=3$, which has one bulk DoF (and four boundary DoF)
and the other having two boundary DoF only. The Maxwell and $U(1)$ Chern-Simons theories
are the simple examples of this. See \cite{Gwak:2015jdo} for more details.}
the massless spin two is not the usual 3D Einstein gravity, but the parity odd cousin of it.
Its Chern-Simons realization is based on $\mathfrak{sl}(2,\mathbb R)_{+1}\oplus \mathfrak{sl}(2,\mathbb R)_{+1}$ 
rather than the usual $\mathfrak{sl}(2,\mathbb R)_{+1}\oplus \mathfrak{sl}(2,\mathbb R)_{-1}$\,.
In other words, the two boundary graviton DoF\footnote{{In AdS space-time, we need to impose an appropriate boundary condition in the time-like conformal boundary 
and this may break gauge symmetries on the boundary. As a result, some boundary values of the AdS gauge fields become additional DoF of the theory. These boundary DoF are particularly important when
the bulk DoF are absent, for instance in 3d gravity and 3d conformal gravity.}
} are relatively ghost breaking the parity symmetry. 
Here, the point that we want to emphasize is that the massless spin two which has $\mu=0=\frac{D-3}2$
needed to be realized in a non-standard manner.
Let us come back to our construction in odd dimensions $D\ge5$, where we found that
the theory of massless and PM spin two fields requires massive spin two fields of higher and higher
$\mu$ but when $\mu$ reaches the point $\frac{D-3}2$ the consistency breaks down.
This may suggest that a non-standard realization of the massive field with $\mu=\frac{D-3}2$
may provide a resolution to this problem.
In $D=3$, the general formula \eqref{ab gen} suggests that our construction allows a consistent solution with three fields:
one PM field  $\phi^{\sst (1)}_{\m\n}=\varphi_{\m\n}$ and two massless fields
$\phi^{\sst (0)}_{\m\n}=h_{\m\n}$ and 
$\phi^{\sst (2)}_{\m\n}=\tilde h_{\m\n}$\,. 
This case is special though,
because the general knowledge of cubic interactions \cite{Joung:2012hz} 
may miss vertices existing only in $D=3$.
The dedicated study of 3D cubic interactions for massless fields
has been carried out in  \cite{Mkrtchyan:2017ixk,Kessel:2018ugi}, while the vertices with PM fields are not studied yet. 

A solution with two copies of massless spin-two fields arises also in all even dimensions. We will study such models in the following. An important difference as compared to the case of $D=3$ is that in even dimensions we will also have two copies of PM fields and that the massless spin-two fields are propagating in dimensions $D\geq 4$.

\subsubsection{Doubled Spectra}\label{sec: doubled spectra}

In the even dimensional analysis of the last section, we could opt for proceeding without 
identification between $\phi^{(\frac{D-2}2)}$ and $\phi^{(\frac{D}2)}$. Even in this case,
we still face the same problem as before since we still need to satisfy
\be
	a^{(\frac{D-2}2)}_+\,b^{(\frac{D-2}2)}_+=-2\,(D-3)\,\l_{hhh}^2\,,
\ee
whereas the sign of $a^{(\frac{D-2}2)}_+\,b^{(\frac{D-2}2)}_+$ 
is given by the relative sign of the kinetic terms of $\phi^{(\frac{D-2}2)}$ and $\phi^{(\frac{D}2)}$
(see Appendix \ref{sec: Cubic vertices} for the details).
Nevertheless, we can proceed with the construction, by choosing 
the kinetic term sign of $\phi^{(\frac{D}2)}$ opposite to $\phi^{(\frac{D-2}2)}$.
Then, we are led to introduce also $\phi^{(\frac{D+2}2)}\,, \phi^{(\frac{D+4}2)}$\,, \ldots,
$\phi^{\sst (D-1)}$ with the same kinetic term signs as $\phi^{(\frac{D}2)}$\,. 
When we reach the last field $\phi^{\sst (D-1)}$\,, which is another massless spin two,
we do not need to introduce the tachyonic $\phi^{\sst (D)}$ because
the transformation between $\phi^{\sst (D-2)}$ and $\phi^{\sst (D-1)}$ alone suffices
to satisfy the symmetry closure up the linearised gauge transformation. 
This is simply due to the symmetry
between \eqref{+ os} and \eqref{- os} under $\m\to D-1-\m$.
In this way, we get exactly doubled spectra:
any field in this theory has its equal mass partner.
Even though the field content is symmetric, the cubic interactions  
responsible for the global symmetry transformations are not. They are
\be
	\phi^{\sst (0)}\!-\!\phi^{\sst (n)}\!-\!\phi^{\sst (n)}\,,\qquad
	\phi^{\sst (1)}\!-\!\phi^{\sst (n)}\!-\!\phi^{\sst (n+1)}\,,
\ee
where $n=0,\ldots, D-1$ and one vertex falls into both classes.
The introduction of another copies of massless and PM fields
$\phi^{\sst (D-1)}$ and $\phi^{\sst (D-2)}$ requests 
to enlarge the vector space of the global symmetry by another copies of 
massless and PM Killing tensors.
Without explicit calculation, we can asses what might be the resulting 
global symmetry.
For that, let us make a schematic analysis where 
$M$ and $K$ denote the original generators forming $\mathfrak{so}(1,D+1)$
and $\tilde M$ and $\tilde K$ the new generators associated with the fields $\phi^{\sst (D-1)}=\tilde \varphi$ and 
$\phi^{\sst (D-2)}=\tilde h$\,.
If we focus on three gauge fields interactions responsible for
the Lie brackets of the global symmetry algebra, then there are
\be
	 h-h-h\,, \qquad h-\varphi-\varphi\,,\qquad h-\tilde\varphi-\tilde\varphi\,,
	\qquad h-\tilde h-\tilde h\,,\qquad \varphi-\tilde\varphi-\tilde h\,.
	\label{cubic inter 4}
\ee
From the above, we can figure out the schematic structure of the Lie brackets as 
\ba
	\,[\![\,M\,,\, M\,]\!]=M\,,\qquad 
	&\,[\![\,M\,,\,  K\,]\!]=K\,,\qquad 
	&\,[\![\,K\,,\,  K\,]\!]=M\,, \nn
	\,[\![\,M\,,\,  \tilde M\,]\!]=\tilde M\,,\qquad 
	&\,[\![\,M\,,\,  \tilde K\,]\!]=\tilde K\,,\qquad 
	\,[\![\,\tilde M\,,\,  K\,]\!]=\tilde K\,,\qquad 
	&\,[\![\,K\,,\,  \tilde K\,]\!]=\tilde M\,,\nn
	\,[\![\,\tilde M\,,\,  \tilde M\,]\!]=M\,,\qquad 
	&\,[\![\,\tilde M\,,\,  \tilde K\,]\!]=K\,,\qquad 
	&\,[\![\,\tilde K\,,\,  \tilde K\,]\!]=M\,.
\ea
The details of the brackets are also under control since 
they are inherited from the cubic interactions \eqref{cubic inter 4}
whose forms are indifferent whether each fields are tilded or not.
In this way, we can conclude
the global symmetry of this theory is $\mathfrak{g}\otimes \mathbb Z_2$\,, where $\mathfrak{g}$ is some real form of $\mathfrak{so}(D+2)$. It is straightforward to check that the algebra splits into two parts with generators,
\be
N=\frac12(M+\tilde M)\,,\quad L=\frac12 (K+\tilde K)\,,\quad \tilde{N}=\frac12(M-\tilde M)\,,\quad \tilde L=\frac12 (K-\tilde K)\,,
\ee
which are mutually commuting:
\ba
\,[\![\,N\,,\, N\,]\!]=N\,,\qquad 
	&\,[\![\,N\,,\, L\,]\!]=L\,,\qquad 
	&\,[\![\,L\,,\, L\,]\!]=N\,, \nn
	\,[\![\,N\,,\,  \tilde N\,]\!]=0\,,\qquad 
	&\,[\![\,N\,,\,  \tilde L\,]\!]=0\,,\qquad 
	\,[\![\,\tilde N\,,\, L\,]\!]=0\,,\qquad 
	&\,[\![\,L\,,\,  \tilde L\,]\!]=0\,,\nn
	\,[\![\,\tilde N\,,\,  \tilde N\,]\!]=\tilde N\,,\qquad 
	&\,[\![\,\tilde N\,,\,  \tilde L\,]\!]=\tilde L\,,\qquad 
	&\,[\![\,\tilde L\,,\,  \tilde L\,]\!]=\tilde N\,.
\ea
The consistency of the $\delta^{\sst [1]}$ transformation by the tilded generators 
can be also achieved thanks to the  $\mathbb Z_2$ structure:
the transformation rules by tilded generators 
are the same as the untilded one upon interchanging one of the field to its 
$\mu\to D-1-\mu$ counterpart.
For this, we would need to include also the cubic interactions,
\be
	\phi^{\sst (D-1)}\!-\!\phi^{\sst (n)}\!-\!\phi^{\sst (D-1-n)}\,,\qquad
	\phi^{\sst (D-2)}\!-\!\phi^{\sst (n)}\!-\!\phi^{\sst (D-2-n)}\,,
\ee
where $n=0,\ldots, D-1$ and some interactions are redundant: they may appear in more than one of the above mentioned classes.
Therefore, this theory seems to pass 
all the consistency requirements considered in this paper,
even though the fields $\phi^{(0\le n\le \frac{D-2}2)}$  and $\phi^{(\frac{D}2\le n\le D)}$ are relatively ghost.
In fact, there could be a trivial relation between this theory and conformal gravity.
Since the global symmetry $\mathfrak{g}\otimes \mathbb Z_2$ can be decomposed into
${\mathfrak{g}}\oplus {\mathfrak{g}}$\,,
one can expect that the theory itself can be also written as a direct sum of 
two mutually non-interacting conformal gravity Lagrangians,
similarly to the Einstein-Hilbert analog considered in \cite{Boulanger:2000rq}.

Let us consider now the $D=3$ case.
As one could check from the general formula \eqref{ab gen}, there is a consistent solution with three fields $h_{\m\n}$, $\varphi_{\m\n}$ and $\tilde h_{\m\n}$. Since  $a^{\sst (1)}_+\,b^{\sst (1)}_+=-2\,\l_{hhh}^2<0$, the third field $\tilde h_{\m\n}$ 
has the opposite kinetic term sign  to $h_{\m\n}$ and $\varphi_{\m\n}$.
Similarly to the even dimensional case that we  discussed just above,
the 3d theory under consideration will have an enlarged global symmetry, 
\mt{\mathfrak{so}(1,3) \subset\hspace{-11pt}+\ {\mathbb R^{1,3}}\,+\hspace{-11pt}\supset \mathfrak{so}(1,3)},
the semi-direct sum between two copies of dS$_3$ isometry and the Abelian PM symmetries.
Since the PM field $\varphi_{\m\n}$ is realized as a two-derivative Lagrangian,
its degrees of freedom (DoF) are composed of 1 bulk and possibly 4 boundary DoF.
Concerning the boundary DoF, we may consider 
the decomposition $4\to 2+2$ where $2$ is the boundary DoF 
contained in the Chern-Simons formulation of 3D Conformal Gravity \cite{Afshar:2011qw}.
However, due to the presence of one bulk DoF,
the 3D theory under consideration is
different from the direct sum of two copies of 3D Conformal Gravity.
It would be interesting to identify the full interacting theory in 3D  that we found up to cubic order in
interaction.

Finally, let us add one more remark on the even dimensional cases.
Even though the direct sum of two conformal gravities should be one of the solutions in this construction,
it is not clear, from the schematic analysis in this section, whether it is the only solution with the given spectrum. 
If there is a room for another option, the resulting theory would be 
a strange cousin of ``doubled conformal gravity", which itself
may admit a non-trivial decomposition into two mutually non-interacting theories\footnote{{It is also natural to guess that the ``multiple conformal gravity'' examples we found can be related to those studied in \cite{Boulanger:2002bt,Boulanger:2001vr}.}}. In four dimensions, it might be also possible to identify the ghost fields with the electric-magnetic dual of the original fields imposing extra constraints, thus establishing a unitary theory.
These speculations lead us to consider a decomposition of parity invariant theory into two parity-violating ones, thus a theory with parity odd couplings.

\section{Relaxing Other Assumptions}
\label{sec: other assum}

\subsection{Parity Violating Theory in $D=4$}
\label{sec: parity odd}

So far, we have implicitly required the invariance under the parity symmetry,
that is, the interaction vertices do not involve any Levi-Civita tensor $\e_{\m_1\cdots \m_D}$.
Now, we examine the consequences of potential parity odd vertices.
We shall restrict the analysis to the dimension $D=4$ because
in dimensions $D\geq 6$ parity odd vertices are absent for symmetric fields, while in $D=5$ the parity-odd self-interactions of a spin-two field requires Chan-Paton factors.
Let us go back to the set up \eqref{V012}.
The vertices $\cV_{hhh}$ and $\cV_{h\varphi\varphi}$ are related to the general covariance, so we will not attempt to modify them.
$\cV_{\varphi\varphi\varphi}$ is the PM self-interaction,
and let us replace it by a parity-odd  analogue, say $\widetilde\cV_{\varphi\varphi\varphi}$.
Then, it would induce a parity-odd first-order deformation on PM gauge transformation.
Due to the antisymmetry of the Levi-Civita tensor,
there is only one possible structure for $\delta^{\sst [1]}_\a\,\varphi_{\m\n}$\,: 
\be
\d^{\sst [1]}_\a \vf_{\m\n}=2\, \tilde\l_{\varphi\varphi\varphi}\,\e_{\r\s\l(\m}\,\partial^\r\a\,
\nabla^\s\,\varphi^{\l}{}_{\n)}\,.\label{PVgt}
\ee
With the above, the first-order deformations of the gauge symmetries  \eqref{h varphi transf} will
be modified to 
\ba
\d_\a^{\sst [1]}\varphi_{\m\n}\eq \frac12\,\l_{hhh} \left[\partial^\r \a\,
(\nabla_\r h_{\m\n}-2\,\nabla_{(\m}h_{\n)\r})+\frac{2\,\L}{3}\,\a\,h_{\m\n}\right]
+2\,\tilde\l_{\varphi\varphi\varphi}\,\e_{\r\s\l(\m}\,\partial^\r\a\,
\nabla^\s\,\varphi^{\l}{}_{\n)}\,,\nn
\d^{\sst [1]}_\a h_{\m\n}\eq 2\,\l_{hhh}\,\partial^\r \a\left(\nabla_\r\varphi_{\m\n}-2\,\nabla_{(\m}\varphi_{\n)\r}\right)\,,
\ea
where we have already used the constraint $\l_{h\varphi\varphi}=\l_{hhh}$ imposed by the Jacobi identity of 
the global symmetry.
Now we can examine again the closure of the above transformations by imposing the Killing condition \eqref{Killing}
and the on-shell conditions \eqref{OS} and \eqref{OPM}.
Note that one can verify that the parity-odd transformation \eqref{PVgt} is compatible with
the on-shell conditions: 
\begin{align}
\Big(\Box-\frac{4\L}{3}\Big)\left[\d^{\sst [1]}_{\bar\a}\varphi_{\m\n}\right]=0\,,\qquad 
\nabla^{\m}\left[\d^{\sst [1]}_{\bar\a}\varphi_{\m\n}\right]=0\,,\qquad 
\bar{g}^{\m\n}\left[\d^{\sst[1]}_{\bar\a}\varphi_{\m\n}\right]=0\,,
\end{align}
which can be shown employing identities from Appendix \ref{sec: details}.
Again the commutators $[\![\bar \xi_1,\bar \xi_2]\!]$ and $[\![\bar \xi,\bar \a]\!]$
are guaranteed to be consistent by the generally covariant interactions $\cV_{hhh}$ and $\cV_{h\varphi\varphi}$,
so we are left with the check of $[\![\bar\a_1,\bar\a_2]\!]_\m= -\frac{\L}3\,\l_{h\varphi\varphi}\,\a_{[1}\partial_\m \a_{2]}$\,.
First, acting the commutators on the on-shell massless spin two $h_{\m\n}$, we find 
\be
	\left(\d^{\sst [1]}_{\bar\a_{1}}\,\d^{\sst [1]}_{\bar\a_{2}}-\d^{\sst [1]}_{\bar\a_{2}}\,\d^{\sst [1]}_{\bar\a_{1}}\right)h_{\m\n}=
	\d^{\sst [1]}_{[\![\bar\a_1,\bar\a_2]\!]} h_{\m\n}
	+\nabla_{(\m}\xi_{\n)}(\bar\a_1,\bar\a_2)
	\sim \d^{\sst [1]}_{[\![\bar\a_1,\bar\a_2]\!]} h_{\m\n}
	\label{commutatorMassless}
\ee
where the parameter of linearised diffeomorphisms is given exactly as $\cB_{\m}$ in \cite{Joung:2014aba}, except that its $\vf$-dependent part is now
\ba
&\xi_{\n}(\bar\a_1,\bar\a_2,\vf)
=8\,\l_{hhh}^2\,\epsilon_{\r\s\l\k}\,\nabla^\r\bar\a_1\,
\nabla^\s\bar\a_2
\,\nabla^\l\,\vf^{\k}{}_{\n}\,.
\ea
Second, acting by the commutator on the PM field, we find
\ba
&\left(\d^{\sst [1]}_{\bar\a_{1}}\,\d^{\sst [1]}_{\bar\a_{2}}-\d^{\sst [1]}_{\bar\a_{2}}\,\d^{\sst [1]}_{\bar\a_{1}}\right)\vf_{\m\n}
\approx \d^{\sst [1]}_{[\![\bar\a_1,\bar\a_2]\!]}\vf_{\m\n}+\left(\l_{hhh}^2-\tilde \l_{\varphi\varphi\varphi}^2\right) \mathcal{C}_{\m\n}\,,
\ea
where $\cC_{\m\n}$ is given in \eqref{cC} and we dropped the terms, proportional to free equations of motion. The full expression is given in Appendix \ref{sec: details}. 
We can see that for $\tilde\lambda_{\varphi\varphi\varphi}=\pm\,\l_{hhh}$
the global symmetry transformations represented by $\d^{\sst [1]}$ close. 
Therefore, the putative theory with a parity-odd $\varphi\!-\!\vf\!-\!\vf$ interaction 
giving rise to \eqref{PVgt} is compatible with unitarity. This happens only for Minkowski signature, with $\epsilon_{\m\n\r\s}\,\epsilon^{\m\n\r\s}=-24<0$\,.
It is worth noting, that the $\d^{\sst [1]}$ transformation
closes only in the case when there is no parity-even self-interaction of PM field (or, $\l_{\varphi\varphi\varphi}=0$).

Now that the theory under consideration passes the admissibility condition, 
we would like to explicitly construct the Lagrangian up to cubic order. Here we encounter a problem: there is no covariant cubic vertex for parity-odd self-interaction of PM spin two field.\footnote{This situation is similar to the well-known example of the action for the particle on the sphere \cite{Witten:1983tw} in presence of magnetic monopole at the centre of the sphere: we can easily construct the equations of motion, but the corresponding interaction term, contracted with the field vanishes, therefore the straightforward candidate Lagrangian vertex is missing. One can try to construct a Wess-Zumino-like interaction term  --- we postpone this to a future work.}
In fact, a parity-odd vertex exists in case if there is another PM field involved (see Appendix \ref{sec: pov}), but it still cannot help with the closure of the algebra, unless the second PM field is a ghost.
This finding is puzzling, but may indicate that the theory we are after cannot be consistently formulated in the variables that we chose. It is somewhat similar to the Metsaev's findings about massless higher-spin fields in four-dimensional flat space \citep{Metsaev:1991mt}: for the consistency one has to include vertices that exist in light-cone, but cannot be written in the manifestly Lorentz-covariant form through Fronsdal fields.

Our results indicate the existence of a unitary theory of massless gravity interacting with PM spin two field (or, simply unitary cousin of Conformal Gravity), that has a global symmetry $\mathfrak{so}(1,5)$. This theory cannot be written in terms of familiar metric variables in the standard Lorentz-covariant manner though. We do not exclude the existence of a more suitable covariant approach,
and  we hope to address it in future works. 
For the moment, we note that 
the $(A)dS$ light-cone approach \cite{Metsaev:1999ui,Metsaev:2000ja,Metsaev:2005ws,Metsaev:2018xip} can be one way to study this problem. 
Another possibility is to study the `amplitudes': the spinor-helicity formulation 
recently developed for $AdS_4$ \cite{Nagaraj:2018nxq} and its suitable generalization might be useful.
It would be worth to remark here that 
the parity-odd cubic self-interaction amplitudes of the massless spin-two (or higher-spin) fields in flat four-dimensional space
exist despite the fact that the corresponding vertices cannot be written in a covariant form in terms of usual Fronsdal variables (see, e.g., \cite{Conde:2016izb}). 
Therefore, we can speculate that the same happens for the parity-odd cubic self-interaction amplitude of the PM field.
To conclude, let us add another speculation that 
the putative formulation of the parity-violating PM Gravity may exhibit the electromagnetic duality symmetry manifestly. 
The manifest duality symmetric formalism is developed for linearised gravity in \cite{Henneaux:2004jw,Deser:2004xt,Julia:2005ze} (see also \cite{Henneaux:2015cda,Henneaux:2016zlu} for a recent generalisation to massless higher-spin fields). This speculation can be further supported by the observation of \cite{Vasiliev:2007yc}
that an analogous symmetry --- $\mathfrak{so}(3,3)$ in $AdS_4$ --- acting on massless fields can be made manifest at the level of equations of motion using duality-symmetric variables.

\subsection{Non-Geometric Theory}
\label{sec: non-grav}

In the analysis above, we assumed that the gravitational interactions of the PM and massive fields are simply given by covariantising the derivatives in the free action, thus viewing it as a matter that couples to gravity universally, obeying the general covariance. In that case, the transformations of these fields with respect to gauge symmetries of the massless spin two are given by the familiar diffeomorphism. 
In fact, the analysis of cubic interactions in (A)dS reveals that there is
yet another two-derivative vertex for the interaction among a massless spin two and two (partially-)massive spin two fields
\cite{Joung:2012hz}. Introduction of this vertex would modify the gauge transformations and therefore violate the general covariance of the PM fields.
From here on, we shall refer to this interaction vertex as ``non-geometric''.
Once the non-geometric coupling is allowed, 
one has to start over the full analysis again.
Therefore, this possibility may lead to a resolution to the obstruction imposed by the admissibility condition.

Now, let us consider 
 the non-geometric $h\!-\!\varphi\!-\!\varphi$ vertex.
It can be written as the current coupling,
\be
\tilde\cV_{h\vf\vf}={\tilde \l}_{h\vf\vf}\,h^{\m\n}\,J_{\m\n}\,,\qquad 
J_{\m\n}=F_{(\m}{}^{\l,\r}\,F_{\n)\l,\r}-\frac14\,\bar g_{\m\n}\,F_{\l\r,\s}\,F^{\l\r,\s}\,,\label{V4}
\ee
where  we introduced the linearised curvature of PM field via
\be
F_{\m\n,\r}=\nabla_\m\varphi_{\n\r}-\nabla_\n\varphi_{\m\r}\,.
\ee
The current $J_{\m\n}$ is conserved in any dimensions, but partially conserved in four dimensions only (conserved and traceless currents are automatically partially conserved). The self-interaction vertex for PM field can actually be written as 
$\mathcal{V}_{\vf\vf\vf}=\l_{\vf\vf\vf}\,\varphi^{\m\n}\,J_{\m\n}$ (see Appendix \ref{sec: Cubic vertices}) and exists only in four dimensions.
The vertex \eqref{V4} is obviously Abelian\footnote{{We remind the reader, that we call a cubic vertex Abelian, if it does not induce deformation of the bracket of the parameters, or equivalently, the global symmetry. That is to say, it can induce $\d^{\sst [1]}$, but the corresponding commutator $\d^{[0]}_{[\e}\d^{\sst [1]}_{\eta]}\chi_i\equiv \d^{\sst [0]}_{[\eta,\e]}\chi_i=0$ vanishes.}} but it induces a deformation of gauge transformation of the PM field.
The transformation of PM field with gauge parameter of massless spin two will be given by
\be
\d_{\xi}^{\sst [1]} \varphi_{\m\n}={\cal L}_{\xi}\,\varphi_{\m\n}+{\cal D}_{\xi}\,\varphi_{\m\n}\,,\label{VarLD}
\ee
where the first term
\be
{\cal L}_{\xi}\,\varphi_{\m\n}=\l_{h\vf\vf}\left(\xi^\r\,\nabla_\r\varphi_{\m\n}+2\,\nabla_{(\m}\xi^\r\,\varphi_{\n)\r}\right),
\ee
is the usual Lie derivative, while the second term is given (up to trivial transformations) by
\be
\mathcal{D}_\xi\, \vf_{\m\n}=-{\tilde \l}_{h\vf\vf}\left(\xi^\l\,F_{\l(\m,\n)}+\frac{D-1}{2\,\L}\,\nabla_{(\m}(\nabla^{\r}\xi^{\l}\,F_{\n)\l,\r})
\right).
\label{D tr}
\ee
Remark that the transformation \eqref{D tr} is not analytic in cosmological constant, therefore does not have a smooth flat space limit. As opposed to the Fradkin-Vasiliev mechanism \cite{Fradkin:1987ks}, where the Lagrangian is non-analytic in cosmological constant while the gauge transformations induced from it are analytic, here the Lagrangian vertex has a smooth flat limit while the gauge transformation, that ensures consistency, does not. 
{The underlying interaction vertex $\tilde \cV_{h\vf\vf}$ was missed in the previous literature,
for instance \cite{Zinoviev:2013hac} and \cite{Joung:2014aba}, thus questioning the applicability of the light-cone \cite{Metsaev:2007rn} and TT classifications \cite{Joung:2012rv} to the off-shell fields.
The implicit assumption of one-derivative transformation lows for two-derivative vertices was the reason for 
the omission of the vertex \eqref{V4} in \cite{Zinoviev:2013hac}. 
The apparent non-cancellations of the trace and divergence terms was the reason of omission in \cite{Joung:2014aba}, where the hidden assumption was that the off-shell completion should be similar to that of massless fields  \cite{Manvelyan:2010jr}.
In fact, we show in Appendix \ref{sec: Cubic vertices} that, thanks to the presence of infrared regulator --- mass or cosmological constant ---
{\it any TT vertex can be promoted to a full off-shell vertex for massive fields too, albeit with slightly more general mechanism than that of massless fields}. 
The special feature of this generic completion is that higher-derivative terms arise in gauge transformation deformations. }

Now in order to test the admissibility condition, we consider the Killing vector $\bar\xi$ and on-shell field $\varphi_{\m\n}$.
Then we get
\begin{align}
{\cal D}_{\bar\xi}\,\varphi_{\m\n}=-\frac{D-1}{4\,\L}\,{\tilde \l}_{h\vf\vf}
\left(\nabla^\r\bar\xi^\s\,\nabla_{(\m|}F_{\r\s,|\n)}+\frac{4\,\L}{D-2}\,\bar\xi^\r\,F_{\r(\m,\,\n)}\right).\label{Dxi}
\end{align}
Interestingly enough, the above precisely cancels with the problematic piece \eqref{cC} 
in the commutator $\delta^{\sst [1]}_{[\bar\a_{1}}\,\delta^{\sst [1]}_{\bar\a_2]}\,\varphi_{\m\n}$
upon the identification,
\be
\tilde\l_{h\vf\vf}=\frac{2\,(D-2)}{D-3}\,\l_{h\vf\vf}\,,\label{l l}
\ee
in arbitrary dimensions and
\be
	\l_{h\vf\vf}\,\tilde\l_{h\vf\vf}=4\,\left(\l_{h\vf\vf}^2+\l^2_{\vf\vf\vf}\right),
	\label{ll ll}
\ee
in four dimensions.\footnote{We take into account \eqref{ComKillA} in this derivation, equivalent to $\bar\xi^\m=[\![\bar\a_1,\bar\a_2]\!]^\m=\frac{2\,\L}{3}\,\l_{h\vf\vf}\,\bar\a_{[1}\partial^\m\bar\a_{2]}$ in four dimensions. Note that the difference between \eqref{l l} and \eqref{ll ll} comes from the PM self-interaction with the coupling constant $\l_{\vf\vf\vf}$ which exists only in four dimensions.}

This is a good news, but not the end of the story: 
as we modified $\delta^{\sst [1]}_{\xi}$,
we have to check also the closure $[\delta^{\sst [1]}_{\bar\xi_1},\delta^{\sst [1]}_{\bar\xi_2}]
=\delta^{\sst [1]}_{[\![\bar\xi_1,\bar \xi_2]\!]}$.
We compute the commutator of two gauge transformations with the vector gauge parameter, and get
\begin{align}
[\delta^{\sst [1]}_{\bar\xi_1},\delta^{\sst [1]}_{\bar\xi_2}]\vf_{\m\n}=\delta^{\sst [1]}_{[\![\bar\xi_1,\bar \xi_2]\!]}\vf_{\m\n}
+\delta^{\sst [1]}_{\bar\zeta(\bar\xi_1,\bar\xi_2)}\vf_{\m\n}\,,
\label{DD com}
\end{align}
where
\begin{align}
[\![\bar\xi_1,\bar \xi_2]\!]^{\m}=\bar{\xi}_1^\r\,\nabla_\r\,\bar \xi_2^\m-\bar{\xi}_2^\r\,\nabla_\r\,\bar \xi_1^\m\,,\label{Com diff}\\
	\bar\zeta_{\m\n\r}(\bar\xi_1,\bar\xi_2)=\bar\xi_{2\,\m}\nabla_{[\n}\bar\xi_{1\,\r]}-\bar\xi_{1\,\m}\nabla_{[\n}\bar\xi_{2\,\r]}\,.
	\label{zeta}
\end{align}
The right hand side of \eqref{DD com} has an extra term $\delta^{\sst [1]}_{\bar\zeta}\vf_{\m\n}$ (whose precise expression is given in the Appendix \ref{sec: details}),
therefore we see that the transformations by vector parameter, deformed by the non-generally-covariant vertex $\tilde \cV_{h\vf\vf}$
are not closed any more.
This problem might be again fixed by introducing another field, say $\o$,
with an interaction $\cV_{\o\vf\vf}$.
If the new field $\o$ has gauge symmetry with a parameter $\zeta$, then the interaction $\cV_{\o\vf\vf}$ 
may also induce a non-trivial transformation $\delta^{\sst [1]}_{\zeta}\varphi_{\m\n}$,
which might be identified with the problematic piece of \eqref{DD com}. For that to happen, there should exist
also a $\o\!-\!h\!-\!h$ interaction and in such a case the global symmetry
$\mathfrak{so}(1,D+1)$ will be enhanced  by the Killing tensor $\bar\zeta$ such  that the new global symmetry
does not include $\mathfrak{so}(1,D+1)$ any more as a subalgebra.
We can notice from \eqref{zeta} that the gauge parameter $\zeta_{\m\n\r}$ has two irreducible components: a fully antisymmetric part and a part with the symmetry of $(2,1)$ hook Young diagram.

A few remarks are in order.
The trace of $\bar\zeta_{\m\n\r}$ is the usual vector commutator of $\bar\xi_1$ and $\bar \xi_2$ given by \eqref{Com diff} (up to terms that vanish for Killing vectors).

In our arbitrary dimensional construction, there are many fields with different mass values involved, and the space of parameters in the theory is large. In fact, all of the fields with $\m=1,2,\dots$ may couple to gravity non-geometrically (see Appendix \ref{sec: Cubic vertices}). We will not attempt the most general analysis here. Instead we note that  
the non-closure term for the $n$-th field can be read off from \eqref{C-} and \eqref{ab gen} and is given by
\begin{align}
\delta^{\sst [1]}_{\bar \a_{[2}}\,\delta^{\sst [1]}_{\bar \a_{1]}}
	\,\f^{\sst (n)}_{\m\n}=
	\left[\delta^{\sst [1]}_{\bar \a_{[2}}\,\delta^{\sst [1]}_{\bar \a_{1]}}
	\,\f^{\sst (n)}_{\m\n}\right]_{\f^{\sst (n-1)}}=\mathcal{L}_{\bar\xi}\,\f^{\sst (n)}_{\m\n}+\cD_{\bar\xi}\,\f^{\sst (n)}_{\m\n}\,,
\end{align}
where $\bar \xi^{\m}$ is given in \eqref{ComKillA}, the $\mathcal{L}_{\bar \xi}\,\f^{\sst (n)}_{\m\n}$ is the diffeomorphism with the parameter $\bar \xi^\m$, and the non-closure term $\cD_{\bar\xi}\,\f^{\sst (n)}_{\m\n}$ is given by
\begin{align}
\cD_{\bar\xi}\,\f^{\sst (n)}_{\m\n}=\frac{2\,\l_{hhh}\,L^2}{D-1-2n}\Big\{
\nabla^\rho\bar\xi^\s\,\nabla_{(\m|}\nabla_{\s}\f^{\sst (n)}_{|\nu)\rho}
	+\frac{D-1}{L^2}\,\bar\xi^\r\,\nabla_{(\m}\f^{\sst (n)}_{\n)\r}
	+\frac{(n-D)(n+1)}{2\,L^2}\,\bar\xi^\r\,\nabla_\r\f^{\sst (n)}_{\m\n})
\Big\}\,.
\end{align}
This transformation indeed corresponds to a non-geometric two-derivative coupling of the type $\f^{\sst (0)}\!-\!\f^{\sst (n)}\!-\!\f^{\sst (n)}$ described in the Appendix \ref{sec: Cubic vertices}. Thus a partial departure from the general covariance is possible by letting all the fields in the theory couple to the massless one in a generally covariant manner, except for one of them. The choice of the field can be arbitrary --- the PM field itself, or one of the massive fields in the ladder considered in previous section. We can stop the ladder at any point and declare the non-closure of the last field as a new transformation coming from the non-geometric coupling to the massless spin two.
For $D=4$ and $\m=n=1$, this expression is what we get from \eqref{cC}, as expected.

In four dimensions, we do not have any other option except to couple the PM field itself to gravity non-geometrically via the coupling \eqref{V4}. This may lead to a unitary theory of massless and PM spin-two fields. We remark here that the gauge field corresponding to the parameter $\zeta_{\m\n\r}$ can be also identified from the definition of $\zeta$ \eqref{zeta} and the Killing conditions for the vector parameters. It follows then that the Killing condition for $\zeta$ should be
\begin{align}
\nabla_{[\m}\bar\zeta_{\n]\l\r}+\nabla_{[\l}\bar\zeta_{\r]\m\n}=0\,,
\end{align}
which indicates that the corresponding gauge field is of rank four, with two pairs of antisymmetric indices,
which are symmetric under the exchange of the pairs. There are two irreducible tensors with such index structure: fully antisymmetric tensor, and the tensor corresponding to window Young diagram $(2,2)$. A more natural language for studying systems with such fields is employing forms, and corresponding deformations of the gauge structure may be related to some sort of higher gauge theory. In four dimensions such fields are expected to be topological, and it might be easier to construct interacting theories there. A priori, we cannot exclude appearance of new propagating degrees of freedom though, e.g. via a non-standard realisation of the ``notivarg'' type \cite{Deser:1980fy}. A more detailed study of such theories will be conducted elsewhere.

\section{Discussion}
\label{sec: discussion}

In this paper, we have revisited the no-go theorem  \cite{Joung:2014aba}  of interacting and unitary theory of PM spin two field
by identifying and relaxing implicit assumptions therein.
We first relaxed the assumption on the field content: the theory includes only massless and PM spin two fields.
It turned out that only a massive spin-two field can cure the obstruction for the closure condition on PM field,
but the problem reappears on the new field, which can be avoided again by adding yet another massive spin two field.
In this way, we are led to introduce more and more massive spin-two fields with specific mass values. When the mass value reaches a certain bound, 
we find that the closure of the algebra is not possible unless we give up  unitarity.
By allowing relative negative signs for kinetic terms in even $D$ dimensions, we find that the minimal possibility is given by the Conformal Gravity Lagrangian. We study also an alternative possibility with doubled spectrum of the Conformal Gravity. 
In odd dimensions $D\ge5$, we do not find any theory compatible with general covariance even relaxing the unitarity.

From the global symmetry point of view, three dimensions is special: the corresponding algebra is not simple, but gives the four dimensional Poincar\'e algebra, $\mathfrak{iso}(1,3)$ in this case. From field-theoretical point of view, the situation in $D=3$ is somewhat similar
to the even dimensional case: by relaxing the unitarity, we close the symmetry with PM fields and doubled massless spin two fields. 
Differently from the even dimensional case, this cannot be a sum of two 3d Chern-Simons Conformal Gravity Lagrangians
because of the mismatch in the global symmetries as well as the bulk degrees of freedom. The existence of two massless ``gravitons'' in this model poses new challenges in constructing the full non-linear theory. 
The corresponding full non-linear theory, if existing, can be the first example of an action for Coloured Gravity with bulk propagation. 
The two gravitons are mutually ghost but not propagating, whereas the propagating PM degree of freedom has a positive norm.	

In the end, we find that relaxing the assumption on the field content alone cannot  overcome the obstruction, so a stronger no-go theorem in arbitrary dimensions is obtained.

We next consider the possibility to relax another assumption of the no-go theorem --- parity invariance of the theory. This leads us to a theory with parity-odd self-interaction of PM field in four dimensions, that passes the admissibility condition. The corresponding parity-odd cubic vertex is missing though, in the metric-like variables. This theory may exist in another formulations and we hope to come back to it in the future.

Another direction we explored here is involving departure from the generally covariant gravitational coupling, introducing another vertex \eqref{V4} of interactions with gravity for PM fields. The special feature of such a coupling is that it induces a gauge transformation that is non-analytic in cosmological constant.
The existence of the cosmological constant as an infrared regulator can be essential in theories involving such couplings, 
and the limit sending the regulator to zero may not be consistent. 
Note that this is a generic feature of interactions involving massive fields (see Appendix \ref{sec: Cubic vertices}).
This feature might be related to 
the subtleties arising in the tensionless limit of the field-theoretical interpretations of String Theory.\footnote{If the flat limit of the non-geometric theory may be defined, it will have larger symmetries as the PM field will split into massless spin-two and massless spin-one fields, with corresponding gauge symmetries. The distinctive feature of such an elusive theory would be the non-diagonal gravitational coupling --- cubic interaction between the two massless spin-two fields and the massless spin-one field, with three derivatives.}
We delegate investigation of the space of theories with such a non-geometric coupling to a future work, but comment here about general features of it.  Once a non-geometric  coupling is allowed, the algebra of global symmetries changes in a way that the $\mathfrak{so}(1,D+1)$ is not a subalgebra of it any more. 
If the matter does not couple to the new mixed-symmetry field $\o$, then it would not experience this enlarged symmetry --- it would see only the standard Einstein-Hilbert gravity and matter coupled to it minimally and possibly to the PM field,\footnote{The coupling to PM field is very constrained for matter fields, see Appendix \ref{sec: scalar coupling}.} at least at the classical level.

One can, of course, try to look at the possible space of theories where both of the above assumptions are relaxed --- a parity-violating theory with non-geometric coupling to gravity. We did not study this possibility, as relaxing each of the assumptions already gives a possibility to construct a unitary theory of PM field, and each of them introduce technical complications on their own: general covariance of the theory with parity-odd couplings ensures the closure of the gravitational sector of the gauge symmetries, while the parity-even structure of the theory with non-geometric couplings allows to hope for having a simple action principle in metric variables. None of these theories is constructed completely in this work though. We hope to come back to each of them in the future.

We should note here that the PM field can take a VEV compatible with Lorentz invariance --- constant times $dS$ metric. This is a solution to the PM equations at least to second order in the fields. Once the PM field has a non-zero VEV, it will have lowest order gauge transformations with vector parameter, induced from diffeomorphism, and the mass terms of massless and PM spin-two fields will be mixed.
We work in a specific basis, where a diagonal quadratic action \eqref{S2} is assumed, hence the PM field has a zero VEV. We cannot exclude that for a certain non-linear theory this basis is singular and one has to give a non-zero VEV to the PM field and therefore consider non-diagonal mass matrix for the spin-two fields involved. We did not pursue this direction here.

Even though we do not find any solution for a PM gravity in odd dimensions, there is a natural candidate to this role in odd dimensions --- Chern-Simons (CS) theory with the algebra $\mathfrak{so}(1,D+1)$ (or the algebra $\mathfrak{so}(2,D)$ for Conformal Gravity). Indeed, the conformal gravity in three dimensions is given by a Chern-Simons action with gauge algebra $\mathfrak{so}(2,3)$ --- conformal algebra in three dimensions. 
The CS conformal gravity in $D=3$ uses the non-propagating PM field, which is different from our set up here. 
A special feature of the CS gravity in $D\geq 5$ is that it does not admit linearisation around maximally symmetric background \cite{Chamseddine:1989nu}, therefore it violates one of the main assumptions in our construction here --- perturbative expansion around $dS_D$ background starting from quadratic action \eqref{S2}. Nevertheless, the number of degrees of freedom for such a theory is computed in arbitrary odd dimensions $D=2n+1$ in \cite{Banados:1995mq} (see also \cite{Banados:1996yj}). We note here that when the gauge algebra is taken to be de Sitter algebra in $D=2n+1$ dimensions with an extra $\mathfrak{u}(1)$ factor,\footnote{To our best knowledge, the degrees of freedom count is not available without the extra $\mathfrak{u}(1)$ factor in the gauge algebra. This extra factor introduces a technical simplification, as explained in \cite{Banados:1995mq,Banados:1996yj}.} $\mathfrak{u}(1)\oplus \mathfrak{so}(1, D)$, the number of degrees of freedom is given by
\begin{align}
\mathcal{N}=2n^3+n^2-3n-1\,.
\end{align}
We speculate that these degrees of freedom are a collection of spin-two fields:
\begin{align}
\mathcal{N}=
\underbrace{(2n^2-n-1)}_{massless}+\underbrace{(2n^2+n-2)}_{PM}+(n-2)\times \underbrace{(2n^2+n-1)}_{massive}\,,
\end{align}
with mass values given by $\m=0,1,\dots,n-1$. 
Interestingly this collection of fields appears in Conformal Gravity in one lower, $D-1=2n$ dimensions. 
Hence, CS theory in $2n+1$ dimensions with $\mathfrak{so}(1,2n+1)$ algebra
may be underlying the putative parity-violating theory of PM gravity in $2n$ dimensions,
through a special dimensional reduction.\footnote{Instead, for the Chern-Simons theory with the gauge algebra $u(1)\oplus \mathfrak{so}(1, D+1)$ or $u(1)\oplus \mathfrak{so}(2, D)$, the corresponding number of degrees of freedom is larger by $D+1=2n+2$, which cannot be a massive spin-two.}
Such a reduction with the choice of algebra $\mathfrak{so}(1,5)$ may suggest the off-shell variables that are better suited to describe the theory found in Section \ref{sec: parity odd} and therefore requires further investigation (see  \cite{Engquist:2007kz,Aros:2013yaa,Albornoz:2018uin} for related works).

\acknowledgments

We thank Th. Basile, J. Bonifacio, V. Lekeu, R. Manvelyan, C. Troessaert and A. Tseytlin for useful discussions and N. Boulanger, D. Francia and E. Skvortsov for useful comments. 
The research of E.J. was supported by the National Research Foundation (Korea) through the grant 2014R1A6A3A04056670. The work of K.M. was partially supported by Alexander von Humboldt Foundation. 
E.J. is also grateful to AEI Potsdam, for the hospitality extended to him during the workshop ``Higher Symmetries and Quantum Gravity''.
E.J. and K.M. thank Erwin Schr\"odinger Institute in Vienna for hospitality during the Workshop ``Higher spins and holography''.
E.J. appreciates also the hospitality of APCTP through the program ``100+4 General Relativity and Beyond" during the completion of this work.

\appendix

\section{Technical details} \label{sec: details} 

\paragraph{Free PM field}

Here we collect some formulas that are used in manipulations with linear expressions involving PM field $\varphi_{\m\n}$ in the gravitational background given by metric $\bar g_{\m\n}$:
\ba
&F_{\m\n,\r}=\nabla_\m \varphi_{\n\r}-\nabla_\n\varphi_{\m\r}\,,
\\
&\mathcal{G}_{\m\n}(\vf)= \left(\Box-\tfrac{2\,D\,\L}{(D-1)(D-2)}\right) \varphi_{\m\n}-2\,\nabla_{(\m}\,\nabla^\rho\,\varphi_{\n)\rho}
+\nabla_{\m}\,\nabla_{\n}\,\varphi^\rho{}_\rho\nn
&-\,\bar g_{\m\n} \left[ \left(\Box-\tfrac{2\,\L}{(D-1)(D-2)}\right) \varphi^\rho{}_\rho-\nabla^\rho\,\nabla^\sigma\,\varphi_{\rho\sigma} \right]\,,\\
&F_{\m\n,\r}+F_{\n\r,\m}+F_{\r\m,\n}=0\,,\label{B1}\\
&\nabla_{\r}F_{\m\n,\s}+\nabla_{\m}F_{\n\r,\s}+\nabla_{\n}F_{\r\m,\s}=0\,,\label{B2}\\
&g^{\m\r}F_{\m\n,\r}
=-\frac{D-1}{2\L}\nabla\cdot\mathcal{G}_\n\,,\\
&\nabla^{\m}F_{\m\n,\r}=\mathcal{G}_{\n\r}-\frac{D-1}{2\L}\nabla_\r\nabla\cdot\mathcal{G}_\n
+\frac{D-1}{2\L}g_{\n\r}\nabla\cdot\nabla\cdot\mathcal{G}\,,\\
&\nabla^\r F_{\m\n,\r}=-\frac{D-1}{2\L}\nabla_\m\nabla\cdot\mathcal{G}_\n
+\frac{D-1}{2\L}\nabla_\n\nabla\cdot\mathcal{G}_\m\,,\\
&(\Box-\frac{2\,(2D-3)\,\L}{(D-1)(D-2)})F_{\m\n,\r}=\nabla_\m\mathcal{G}_{\n\r}-\nabla_\n\mathcal{G}_{\m\r}
-\frac{D-1}{2\L}\nabla_\m\nabla_\r\nabla\cdot\mathcal{G}_\n
+\frac{D-1}{2\L}\nabla_\n\nabla_\r\nabla\cdot\mathcal{G}_\m\,.
\ea
Here $\mathcal{G}_{\m\n}$  is the same as $\mathcal{G}_{\m\n}^{\sst (\frac{2\L}{D-1})}$ defined in \eqref{Gm}.
 The free equation of motion of the PM field is $\mathcal{G}_{\m\n}=0$, therefore, anything that is on-shell zero 
 can be expressed through $\mathcal{G}_{\m\n}$. Note that some expressions are on-shell zero, but cannot be written through equations of motion analytically in cosmological constant.

\paragraph{Killing conditions}

The Killing parameters satisfy the following equations:
\begin{align}
\nabla_{\m}\bar{\xi}_\n+\nabla_\n\bar\xi_\m=0\,,\qquad \nabla_{\m}\,\partial_\n \bar\a+\frac{2\,\L}{(D-1)(D-2)}g_{\m\n}\bar\a=0\,,\\
\nabla_{\m}\nabla_{\n}\,\bar\xi_{\r}+\frac{2\,\L}{(D-1)(D-2)}(g_{\m\n}\,\bar\xi_{\r}-g_{\m\r}\,\bar\xi_\n)=0\,.
\end{align}

\paragraph{PM commutator in parity-odd theory}

In the parity-violating theory of Section \ref{sec: parity odd}, the commutators of the PM gauge transformations acting on the PM field, are given as
\ba
&[\d^{\sst [1]}_{\bar\a_2},\d^{\sst [1]}_{\bar\a_1}]\vf_{\m\n}=4(-\tilde{\l}_{\vf\vf\vf}^2+\l_{h\vf\vf}^2)[\partial_\r\bar\a_{[1}\partial_\s\bar\a_{2]}\nabla_{(\m}\nabla^{\r}\vf^{\s}_{\,\,\,\n)}
-\L\bar\a_{[1}\partial^\r\bar\a_{2]}(\nabla_{(\m}\vf_{\n)\r}-\nabla_{\r}\vf_{\m\n})]\quad\nn
&+4\tilde{\l}_{\vf\vf\vf}^2\nabla_{(\m}[(\bar{\a_{[1}}\nabla^\r\bar{\a_{2]}})(\cG_{\n)\r}(\vf)-g_{\n)\r} \cG_{\s}{}^{\s}(\vf))]
-\frac{6\tilde\l_{\vf\vf\vf}^2}{\L}\nabla_{(\m}\bar\a_{[1}\nabla^\b\bar\a_{2]}
\,\nabla_{\n)}\nabla_\s\,\cG^{\s\g}(\vf)\nn
&+2\tilde\l_{\vf\vf\vf}^2g_{\m\n}\bar\a_{[1}\nabla_\a\bar\a_{2]}\nabla_\b \cG^{\a\b}(\vf)-2\tilde\l_{\vf\vf\vf}^2\bar\a_{[1}\nabla_{(\m}\bar\a_{2]}\nabla_\a \cG^{\a}{}_{\n )}(\vf)\nn
&+\tilde\l_{\vf\vf\vf}\,\l_{h\vf\vf}\,\epsilon_{\a\b\g(\m}\nabla^\a\bar\a_1\nabla^\b\bar\a_2\,
\cG^{\sst(0)}{}^{\g}{}_{\n)}(h)\,\nn
&=4(-\tilde{\l}_{\vf\vf\vf}^2+\l_{h\vf\vf}^2)\,\mathcal{C}_{\m\n}+\cO(\cG_{\m\n}^{\sst (0)}(h),\,\cG_{\m\n}(\vf)).
\ea
The last expression reduces to the on-shell closure of the algebra discussed in Section \ref{sec: parity odd} for $\tilde\l_{\vf\vf\vf}=\l_{h\vf\vf}$.

\paragraph{Commutator in the non-geometric theory}

Once we allow for the additional vertex \eqref{V4} of non-geometric coupling of the PM field to gravity, 
the gauge transformations of the PM field with the vector parameter of massless spin-two field is modified to
\begin{gather}
\d^{\sst [1]}_\xi \vf_{\m\n}=
\l_{h\vf\vf}(\xi_\r\nabla^\r\vf_{\m\n}+2\nabla_{(\m}\xi^\r\vf_{\n)\r})
-\frac{D-1}{4\L}\tilde \l_{h\vf\vf}(\nabla^\r\xi^\s\nabla_{(\m|} F_{\r\s|\n)}+\frac{4\L}{D-2} \xi^\r F_{\r(\m\n)})\,.
\end{gather}
The commutator of two such transformations is given as in \eqref{DD com}, where
\begin{gather}
\d^{\sst [1]}_{\zeta}\vf_{\m\n}=
\frac{(D-1)^2}{8\L(D-2)} \tilde \l_{h\vf\vf}^2 \zeta^{\l\t\r}\nabla_\m\nabla^\l F_{\r\t\n}
+\frac{(2D-3)(D-1)}{8\L(D-2)}\tilde \l_{h\vf\vf}^2\nabla_\m \zeta^{\l\r\t}\nabla_\l F_{\t\r\n}\nonumber \\
-\frac{(D-1)^2}{8\L(D-2)} \tilde \l_{h\vf\vf}^2 \nabla_\m \zeta^{\l\r\t}\nabla_\n F_{\t\r\l}
+\frac{(D-1)(2\l_{h\vf\vf}-3\tilde \l_{h\vf\vf})}{8\L}\tilde \l_{h\vf\vf}\nabla^\l \zeta_{\t}^{\ \t\r}\nabla_\m F_{\r\l\n}\nonumber \\
+\frac{(D^2-2D+3)(D-1) }{16\L(d-2)}\tilde \l_{h\vf\vf}^2\nabla_\t \zeta^{\t\l\r}\nabla_\m F_{\r\l\n}
-\frac{(2D-3)}{4(D-2)^2}\tilde \l_{h\vf\vf}^2\zeta_{\m\l\r}F^{\r\l}_{\ \ \ \n}\nonumber \\
-\frac{(D^2-3D+3)}{2(D-2)^2}\tilde \l_{h\vf\vf}^2\zeta^{\l\r}_{\ \ \m}F_{\n\l\r}
-\frac{(2D^2-5D+4)}{2(D-2)^2}\tilde \l_{h\vf\vf}^2\zeta^{\l\r}_{\ \ \m}F_{\l\r\n}\nonumber \\
-\frac{(2D-1)}{4(D-2)^2}\tilde \l_{h\vf\vf}^2g_{\m\n}\zeta^{\l\t\r}F_{\t\r\l}
+\frac{3\l_{h\vf\vf}(D-1)-\tilde \l_{h\vf\vf} (2D-1)}{2(D-2)}\tilde \l_{h\vf\vf}\zeta_{\l}^{\ \l\r}F_{\m\r\n}.
\end{gather}
In four dimensions, we get
\begin{gather}
\d^{\sst [1]}_{\zeta}\vf_{\m\n}=
\frac{9}{16\L} \tilde \l_{h\vf\vf}^2 \zeta^{\l\t\r}\nabla_\m\nabla^\l F_{\r\t\n}
+\frac{15}{16\L}\tilde \l_{h\vf\vf}^2\nabla_\m \zeta^{\l\r\t}\nabla_\l F_{\t\r\n}\nonumber \\
-\frac{9}{16\L} \tilde \l_{h\vf\vf}^2 \nabla_\m \zeta^{\l\r\t}\nabla_\n F_{\t\r\l}
+\frac{3(2\l_{h\vf\vf}-3\tilde \l_{h\vf\vf})}{8\L}\tilde \l_{h\vf\vf}\nabla^\l \zeta_{\t}^{\ \t\r}\nabla_\m F_{\r\l\n}\nonumber \\
+\frac{33}{32\L}\tilde \l_{h\vf\vf}^2\nabla_\t \zeta^{\t\l\r}\nabla_\m F_{\r\l\n}
-\frac{5}{16}\tilde \l_{h\vf\vf}^2\zeta_{\m\l\r}F^{\r\l}_{\ \ \ \n}\nonumber \\
-\frac{7}{8}\tilde \l_{h\vf\vf}^2\zeta^{\l\r}_{\ \ \m}F_{\n\l\r}
-2\tilde \l_{h\vf\vf}^2\zeta^{\l\r}_{\ \ \m}F_{\l\r\n}\nonumber \\
-\frac{7}{16}\tilde \l_{h\vf\vf}^2g_{\m\n}\zeta^{\l\t\r}F_{\t\r\l}
+\frac{9\l_{h\vf\vf}-7\tilde \l_{h\vf\vf}}{4}\tilde \l_{h\vf\vf}\zeta_{\l}^{\ \l\r}F_{\m\r\n}\,.
\end{gather}
The free Lagrangian of the fourth rank gauge field with the parameter $\zeta_{\m\n\r}$ and its interactions with PM and massless spin-two fields will be identified elsewhere.

\section{Cubic vertices of spin-two fields}
\label{sec: Cubic vertices}

Here we list cubic vertices involving massless, PM and massive spin two fields in arbitrary space-time dimensions, relevant to our work. We first discuss subtle aspects of general relevance to the interacting theories involving (partially-)massless and massive fields. These are the freedom of field redefinitions and deformations of gauge transformations induced by interactions.

\subsection{Field redefinitions}
When writing an ansatz for the cubic interaction one has to take into account the freedom of field redefinitions. If we redefine a field by
\begin{gather}
	\f_i\to \f_i+c^{jk}_{i}\f_j\f_k\,,
\end{gather}
where $c^{jk}{}_i$ is an operator involving derivatives and tensor contractions in general, then the free action transforms to another one with cubic terms that are proportional to the free equations of motion. These type of cubic interaction terms can be always removed by a field redefinition. This allows to fix a basis for the cubic interaction terms. In case of massless fields, this is done by removing all contractions between derivatives \cite{Manvelyan:2010wp,Manvelyan:2010jr}, which brings these vertices into so-called Metsaev basis \cite{Metsaev:2005ar} (see also \cite{Joung:2012fv,Kessel:2018ugi}). The same can be done for massive fields: using the wave equations of the massive field, one can remove all derivative contractions from the cubic vertices (note that this is not possible in higher order vertices, due to existence of derivative contractions -- Mandelstam variables -- that are not related to d'Alembertian operators acting on separate fields up to partial integrations). In fact, that does not fix all the field redefinition freedom in the case of massive fields. One crucial difference between massless and massive field equations is that the divergence and the trace of the field are not gauge degrees of freedom in the massive case, but can be removed due to second class constraints. This means, that the divergence and the trace can be expressed in terms of the proper field equations of massive fields. In this paper we will deal with the spin-two examples, but let us first illustrate this using massive vector, aka Proca field.

\subsubsection*{Proca field}

The free equations for the Proca field in flat space are given as
\begin{align}
G_{\m}=\Box\,A_\m-\partial_{\m}\,\partial^\n A_\n-m^2 A_\m=0\,,
\end{align}
which has a consequence for $m^2\neq 0$,
\begin{align}
\partial^\m A_{\m}=-\frac1{m^2}\,\partial^{\m}G_{\m}=0\,.\label{ProcaDiv}
\end{align}
This allows to use field redefinitions to remove all the terms in the cubic vertex, proportional to the divergence of the Proca field. In order to show this, we start from the free Lagrangian for Proca field,
\begin{align}
\mathcal{L}_0(A)=-\frac14\, F_{\m\n}F^{\m\n}-\frac12\,m^2\,A_\m\,A^\m\,,\label{FreeProca}
\end{align}
where $F_{\m\n}=\partial_{\m}A_{\n}-\partial_\n A_\m\,,$ and consider a field redefinition of the following kind:
\begin{align}
A_\m\rightarrow \tilde A_\m=A_\m+g\,\frac1{m^2}\,\partial_\m\, f(A, \f_i)\,,\label{FieldRedefinitionProca}
\end{align}
where $g$ can be identified with a small parameter of perturbative expansion in fields. Here $f(A, \f_i)$ is a polynomial function of the Proca field and all other fields $\f_i$ ($i$ is a collective index, which denotes different species of fields) in the theory under consideration, that does not contain constant and linear terms in the polynomial expansion. This field redefinition, plugged back into the free Lagrangian for Proca field \eqref{FreeProca}, gives the following new Lagrangian (we ignore boundary terms):
\begin{align}
\mathcal{L}_0(\tilde A_\m)=\mathcal{L}_0(A)+g\,f(A, \f_i)\,\partial^\m A_\m+g^2\,\frac1{2\,m^2}\,f(A, \f_i)\,\Box\,f(A, \f_i)\,.\label{FreeProcaRedefined}
\end{align} 
The Lagrangian \eqref{FreeProcaRedefined} is a rewriting of free Proca Lagrangian \eqref{FreeProca} in different variables. One can always use proper field redefinitions of the kind \eqref{FieldRedefinitionProca} to generate any interaction terms, proportional to divergences at a given order, and possibly terms of higher order. In particular, any term proportional to the divergence of the Proca field can be removed from the Lagrangian at any given order. In general, this procedure does not only remove the interactions containing divergence of the Proca field, but also adds higher order terms. This is the case also with the field redefinition that brings the cubic vertices of massless fields into the Metsaev basis: this redefinition may generate higher order terms. While working in a given perturbative scheme up to certain order in fields,
 one can always redefine fields in such a way to remove the terms proportional to divergences from interactions up to the given order. 
 Moreover, we can fix  the field redefinition freedom uniquely by removing the terms proportional to $\Box A_\m$ and $\partial^\m A_\m$. With the assumption of manifest Lorentz covariance of the Lagrangian, this fixes the field redefinition freedom completely. Note that for the massless fields we can only get rid of $\Box A_\m$, while the $\partial^\m A_\m$ terms will be fixed by gauge invariance with no extra freedom. Another interesting feature is that the massless limit might be smooth in one form of the action but not in another. The field redefinition itself is not necessarily analytic in mass.
 
\subsubsection*{Massive spin-two} 
 
The argument, given for the Proca field, generalises to the case of fields with spin two and higher. In fact, for massive fields with spin greater than one and a general value of mass, not only the divergence but also trace can be removed by field redefinitions. The reason is that the trace, in the same way as the divergence, can be locally expressed in terms of the free equation of motion.
We will illustrate this for massive spin two fields, which are of interest in this paper. We will work in a background with arbitrary cosmological constant.
For a spin-two field with mass $m$ we have:
\begin{gather}
\nabla^\m\f_{\m\n}-\nabla_\n\,\f=-\frac{1}{m^2}\nabla^\m \mathcal{G}_{\m\n}(\f)\,,\label{Div} \\
\f=\frac1{m^2\,((D-1)\,m^2-2\L)}((D-2)\,\nabla^\m\nabla^\n \mathcal{G}(\f)_{\m\n}+m^2\,\bar g^{\m\n}\,\mathcal{G}_{\m\n}(\f))\,,\label{Tr}
\end{gather}
where $\mathcal{G}_{\m\n}(\f)$ was given explicitly in \eqref{Gm}, $\f=\bar{g}^{\m\n}\,\f_{\m\n}$ and $\bar g_{\m\n}$ is the background $(A)dS$ or flat metric. It is obvious that one can use the first constraint for any massive fields $m^2\neq 0$, while the second one can be used for all mass values except for massless and PM fields: $0\neq m^2\neq \frac{2\,\L}{D-1}$. In flat space, $\L=0$, there are no partially massless fields, and there is only one special value: $m^2=0$.
We first consider the generic massive fields and then comment  on the special PM value of mass.

For the general massive field, there is always a field redefinition, which removes both divergences and traces of the field in the vertices, up to higher order terms. In order to remove an interaction term $\f\,f(\f_i)$, containing the trace $\f$,
\begin{align}
\mathcal{L}(\f)=\frac12 \f^{\m\n}\,\mathcal{G}_{\m\n}(\f)+g\,\f\,f(\f_i)+\cO(g^2)\,,
\end{align}
where $f(\f_i)$ is a polynomial function of all the fields in the theory, we can employ a field redefinition:
\begin{align}
\tilde\f_{\m\n}=\f_{\m\n}+g\,\frac1{m^2\,((D-1)\,m^2-2\L)}((D-2)\,\nabla_\m\nabla_\n +m^2\,\bar g_{\m\n})\,f(\f_i)\,,
\end{align}
to arrive at the action,
\begin{align}
\mathcal{L}(\tilde{\f})=\frac12 \tilde\f^{\m\n}\,\mathcal{G}_{\m\n}(\tilde\f)+\cO(g^2)\,.\label{Og2}
\end{align}
In general, after the redefinition, the new Lagrangian will contain new terms of order $g^2$, and possibly higher. In the PM limit, the field redefinition diverges, while the residue becomes the gauge transformation for the PM field, in the same way as for the Proca field.
In the same way, in the Lagrangian with an interaction term $A^{\n}(\f_i)\,\nabla^{\m}\f_{\m\n}$,
\begin{align}
\mathcal{L}(\f)=\frac12 \f^{\m\n}\,\mathcal{G}_{\m\n}(\f)+g\,A^{\n}(\f_i)\,\nabla^{\m}\f_{\m\n}+O(g^2)\,,
\end{align}
one can implement St\"uckelberg-type non-linear field redefinition
\begin{align}
\tilde{\f}_{\m\n}=\f_{\m\n}+g\,\frac1{m^2}\nabla_{(\m}\,A_{\n)}(\f_i)-g\,\frac1{m^2\,((D-1)\,m^2-2\L)}((D-2)\,\nabla_\m\nabla_\n +m^2\,\bar g_{\m\n})\,\nabla^\r A_\r(\f_i)\,,
\end{align}
to end up with an action of the form \eqref{Og2}. Again, in the limit $m^2\rightarrow 0$, the redefinition diverges, while the residue becomes a gauge transformation for the massless field. In a sense, the redundancy of the divergence and trace terms, that are fixed in the (partially-)massless case by gauge symmetry, is fixed by field redefinition freedom in the massive case, where the field redefinition has the same structure as the linearised gauge transformations of the (partially-)massless field.
For example, in the cubic interactions of massless fields, all the terms with divergences and traces get fixed by gauge invariance, therefore one can concentrate on the TT part only for classification purposes.

For partially-massless spin-two field, the divergence can be removed by field redefinition, but not the trace. Instead, there is enough gauge freedom to gauge-fix the trace to zero and consider TT vertex for classification purposes, while trace terms will be fixed by gauge invariance. In case of the massive fields, TT terms define the vertex fully as both divergences and traces can be removed by field redefinitions.

This implies that the TT terms for the massive field is sufficient to define interactions, but not for gauge fields. Removing divergences and traces of massive fields from interaction vertices gives the simplest possible form for the Lagrangian, while the gauge transformations may be simpler in another field frame.

\subsection{Gauge transformation deformations}

There is another technical aspect, related to the second class constraints \eqref{ProcaDiv}, \eqref{Div}, \eqref{Tr}, that we are going to use in this work. While studying constraints imposed by gauge symmetries in possible interactions, we are usually led to construct a part of any vertex that is given by TT fields, then complete it to full off-shell vertex. The TT vertex is gauge invariant only up to terms that are divergences, traces and Klein-Gordon operators acting on all the fields. It is not generally known whether there exists a completion to full off-shell vertex for a given TT vertex. For the massless symmetric fields this is proven by construction\footnote{In reference \cite{Francia:2016weg} both flat and $(A)dS$ vertices were constructed for Maxwell-like fields carrying either reducible or irreducible massless representations. The construction of vertices in flat space given there is complete both for irreducible and for reducible fields. The full off-shell vertex for irreducible Maxwell-like fields in $(A)dS$ space is also complete, as well as the corresponding generating functions for $(A)dS$ irreducible Maxwell-like and Fronsdal off-shell vertices. The full off-shell vertex for reducible Maxwell-like fields in $(A)dS$ space, however, is incomplete. Its full form will be given in future work \cite{InPreparation}.} \cite{Manvelyan:2010jr,Francia:2016weg}. 
We will give a simple argument here that the off-shell completion of the TT vertices is always possible also in the presence of massive fields.
When relaxing the TT conditions, in the variation of TT vertex one encounters terms that contain divergences and traces of the fields that are supposed to sum up to zero for consistency. In all known examples, divergences and traces of massless fields conspire to cancel, if we add certain non-TT terms into the vertex, allowing for off-shell cubic vertices, while for the massive fields they do not have to. The reason is that, as opposed to massless fields, the divergence and trace of the massive field can be treated as terms proportional to equations of motion. Therefore, the would-be obstruction terms are not obstructing any more, as they can be compensated by a local field redefinition,
 which induces deformation of gauge transformations.

There is a non-trivial point here though. The divergence and trace terms are related to the divergence \eqref{Div} and trace \eqref{Tr} of the equation of motion, divided by mass of the field. This means that whenever a divergence and trace term is treated as equation of motion in the variation of the vertex and compensated by a deformation of gauge transformation,
the induced gauge transformations for these fields will involve non-analyticity in mass, therefore will not admit a consistent massless limit. In a particular case of interest in this paper, for the PM field or other massive fields with mass given through cosmological constant, the massless limit coincides with flat space limit and may be inconsistent with the transformations induced in these type of interactions. Indeed, we encounter such an interaction, given through \eqref{V4}, with gauge transformation deformation given by \eqref{D tr}.

\subsection{Cubic interactions for Massless and PM fields}
\label{sec: EH}

In the eq. \eqref{V012}, we started from three vertices: $\cV_{hhh}, \cV_{h\vf\vf}$ and $\cV_{\vf\vf\vf}$. The first of them is the cubic vertex of Einstein-Hilbert action, the TT part of which can be given in the following simplest form\footnote{The full off-shell vertex in flat space in this simplest form is given in \cite{Manvelyan:2010wp}, while the full off-shell vertex in $(A)dS$ is given in \cite{Manvelyan:2012ww}, in a different field frame.}:
\begin{align}
\cV_{hhh}^{\rm\st TT}=\frac12 h^{\a\b}\nabla_\a\nabla_\b h^{\m\n}\,h_{\m\n}+h^{\a\m}\nabla_{\a}h^{\n\b}\nabla_\b h_{\m\n}+\frac{2\,\L}{3(D-2)} \, h_\m{}^\n h_{\n\r} h^{\r\m}\,.
\end{align}
The second vertex, $\cV_{h\vf\vf}$, is the vertex of minimal coupling to gravity and can be extracted from free PM action with full covariant derivatives. Its TT part is given by
\begin{align}
\cV_{h\vf\vf}^{\rm\st TT}=-\frac12\, h^{\m\n}\nabla_{\m}\nabla_\n \vf^{\l\r}\,\vf_{\l\r}+2\,\nabla_{\l} h^{\m\n}\nabla_{\m}\vf_{\n\r}\,\vf^{\l\r}+3\,\nabla_\m\nabla_\n h_{\l\r}\,\vf^{\l\r}\,\vf^{\m\n}\nn
+3\,\nabla_{\l}h_{\n\r}\,\vf^{\m\n}\,\nabla_\m\vf^{\l\r}+\frac{2\,\L}{D-2}\,h^{\m\n}\,\vf_{\n\l}\,\vf^\l{}_\m\,,
\end{align}
and is a linear combination of \eqref{PM M M} and \eqref{AbelianPMMM} for $\m_1=0$ and $\m_2=1$. The third vertex, $\cV_{\vf\vf\vf}$, is the cubic self-interaction of PM field, employed by Conformal Gravity in four dimensions. Its explicit expressions is given in \cite{Joung:2014aba}. Here we provide a simpler expression for it, which is equivalent to the one there, up to field redefinitions
\begin{align}
\cV_{\vf\vf\vf}=\vf^{\m\n}\Big(F_{\m}{}^{\r,\s}F_{\n\r,\s}-\frac14 \bar g_{\m\n} F_{\r\s,\l}F^{\r\s,\l}\Big)\,.
\end{align}
Note, that this vertex is the full off-shell one.
One can notice the similarity with the vertex \eqref{V4} of the non-geometric coupling between PM and massless spin-two.

\subsection{Cubic interactions for PM field and two massive fields}

We derive the cubic interaction of a PM field with two arbitrary massive fields and its corresponding gauge transformation. Here we study only vertices that have two derivatives and induce gauge transformation deformations found in Section \ref{sec: enlarge FC}. Starting from a general ansatz for the cubic interactions and imposing gauge invariance we are left with four vertices for generic values of masses. These vertices do not induce gauge transformation deformations, and can be expressed in terms of PM curvature:
\begin{gather}
F^{\l\m,\r}{\f_1}_{\m\n}\nabla^\n{\f_2}_{\l\r}\,,\quad
\nabla_\m F^{\n\l,\r}{\f_1}_{\l\r}{\f_2}_{\n}{}^{\m}\,,\quad
F^{\m\l,\n}\nabla_\n{\f_1}_{\l\r}{\f_2}_\m{}^\r\,,\quad
F^{\m\l,\n}\nabla_\m{\f_1}_{\l\r}{\f_2}_\n{}^\r\,,\label{AbelianPMMM}
\end{gather}
where the last two vertices are equivalent up to a factor when $\f_1{}_{\m\n}\equiv\f_2{}_{\m\n}$.
For general mass values, these are all the vertices.
For a special case of $\m_2=\m_1+1$, there is one more vertex, inducing gauge transformations. This special vertex can be written in the form:
\begin{gather}
\mathcal{V}_{1,\,\m_1,\,\m_1+1}=\,\l_{1,\m_1,\m_1+1}\,(\vf^{\m\n}\nabla_\m\nabla_\n {\f_{1}}^{\l\r}{\f_{2}}_{\l\r}
-\vf^{\m\n}\nabla_\n\nabla_\l{\f_1}^{\l\r}{\f_2}_{\m\r}
+\vf^{\m\n}\nabla_\n{\f_1}_{\m\l}\nabla_\r{\f_2}^{\r\l}\nonumber \\
-\frac{4\L}{(D-1)(D-2)}\vf^{\m\n}{\f_{1}}_{\n\l}{\f_{2}}^{\l}_{\m}
-\frac{1}{2}\vf^{\m\n}\nabla_\m{\f_1}_{\n\l}\nabla^\l{\f_2}^\r_\r
+\frac{1}{2}\vf^{\m\n}\nabla_\m\nabla^\l{\f_1}_{\n\l}{\f_2}^\r_\r\nonumber\\
-\frac{1}{2}\nabla^\m\vf^{\n\l}\nabla_\l{\f_1}^\r_\r{\f_2}_{\m\n}
-\vf^{\m\n}\nabla_\n{\f_1}^\r_\r\nabla^\l{\f_2}_{\l\m}
-\vf^{\m\n}\nabla_\m\nabla_\n{\f_1}^\r_\r{\f_2}^\l_\l\nonumber\\
-\frac{(\m_1-2)\L}{(D-1)(D-2)}\vf^{\m\n}{\f_1}^\l_\l{\f_2}_{\m\n}
+\frac{(\m_1+4-D)\L}{(D-1)(D-2)}\vf^{\m\n}{\f_1}_{\m\n}{\f_2}^\l_\l
)+\mathcal{O}(\vf^\r{}_\r).\label{PM M M}
\end{gather}
and induces the gauge transformations ($\m_2\equiv \m_1+1$):
\begin{gather}
\delta^{\sst [1]}_{\a}{\f_1}_{\m\n}=\frac1{D-\m_2}\,\l_{1,\m_1,\m_2}\,(\frac{D-\m_2}{2}\nabla^\r\a\nabla_\r{\f_2}_{\m\n}
-\nabla^\r\a\nabla_{(\m}{\f_2}_{\n)\r}
-\frac{(D-2-\m_2)(D+1-\m_2)\L}{(D-1)(D-2)}\a{\f_2}_{\m\n}) \nonumber \\
\delta^{\sst [1]}_{\a}{\f_2}_{\m\n}=\frac1{\m_2}\,\l_{1,\m_1,\m_2}\,
(\frac{\m_2}{2}\nabla^\r\a\nabla_\r{\f_1}_{\m\n}
-\nabla^\r\a\nabla_{(\m}{\f_1}_{\n)\r}
-\frac{(\m_2-2)(\m_2+1)\L}{(D-1)(D-2)}\a{\f_1}_{\m\n})
\end{gather}
Note, that the form of the cubic interaction is not minimal, and could be simplified by field redefinitions to a form:
\begin{align}
\mathcal{V}_{1,\,\m_1,\,\m_1+1}=\vf^{\m\n}\nabla_\m\nabla_\n {\f_{1}}^{\l\r}{\f_{2}}_{\l\r}
-\frac{4\L}{(D-1)(D-2)}\vf^{\m\n}{\f_{1}}_{\n\l}{\f_{2}}_{\m}{}^{\l}
+\mathcal{O}(\vf^\r{}_\r)\,,
\end{align}
up to an overall factor.
In this field frame, the gauge transformation is modified and includes also four derivative terms. Nevertheless, the global part of the transformations cannot be modified by field redefinitions, therefore our analysis of the algebra closure is independent on the field frame. 

It is worth to note that all of the information about the deformations of gauge transformations induced by non-abelian interactions of PM and massive spin two fields was possible to derive from on-shell conditions only. We can see now that these transformations indeed are those induced from cubic vertices and write down the explicit form of these vertices. Using this identification, we deduce
\ba
a_+ &= -\frac1{D-\m_\chi-1}\, \l_{\chi+}\,,\hspace{65pt} b_+ &=-\frac1{\m_\chi+1}\,\l_{\chi+}\,,\nn
a_- &= -\frac1{\m_\chi}\, \l_{\chi-}\,,\hspace{85pt}  b_- &=-\frac1{D-\m_\chi}\,\l_{\chi-}\,,\nn
a_+\,b_+&=\frac1{(\m_\chi+1)(D-\m_\chi-1)}\,\l_{\chi+}^2\,,\qquad a_-\,b_-&=\frac1{\m_\chi\,(D-\m_\chi)}\,\l_{\chi-}^2\,.
\ea
In the last expressions of $a_\pm b_\pm$, the overall sign will change if we consider the two fields with masses $\m_\chi$ and $\m_\chi \pm 1$ to have opposite-sign kinetic terms.

\subsection{Cubic interactions of a massless and two massive spin two fields}

In order for a cubic vertex inducing deformations of gauge transformations to exist, the two massive fields should have the same mass. Here we take the two massive fields to be actually identical -- this is the case relevant for us in this work.
There are three two-derivative vertices in this case. Their TT parts are given as
\begin{gather}
\cV_1^{\rm\st TT}=h^{\m\n}\,\nabla_\m\,\f^{\l\r}\,\nabla_\n\,\f_{\l\r}+\frac{4\L}{(D-1)(D-2)}h^{\m\n}\f_{\n\l}\f^\l{}_\m\,,\\
\cV_2^{\rm\st TT}= h^{\m\n}\,\nabla_\m\,\f^{\l\r}\,\nabla_\l\,\f_{\n\r}+\frac{(D+1)\L}{(D-1)(D-2)}h^{\m\n}\f_{\n\l}\f^\l{}_\m\,,\\
\cV_3=\Big(\nabla_\m\nabla_\n h_{\l\r}+\nabla_\l\nabla_\r h_{\m\n}-\nabla_\m\nabla_\l h_{\n\r}-\nabla_\n\nabla_\r h_{\m\l}\nn
-\frac{4\L}{(D-1)(D-2)}(g_{\m\l}h_{\n\r}-g_{\m\n}h_{\l\r})\Big)\,\f^{\m\n}\,\f^{\l\r}\,,
\end{gather}
The first two induce deformations of gauge transformations, while the third one contains the linearised curvature of the massless spin-two field, therefore does not induce a deformation of gauge transformations for the fields involved. Only the combination $\cV=-\frac12\,\cV_1+2\,\cV_2+\a \cV_3$ ($\a$ is arbitrary) corresponds to the minimal ``geometric'' coupling to Gravity, which induces a gauge transformation in the form of diffeomorphism for the massive field.

\subsection{Parity-odd cubic (self-)interaction for PM fields}
\label{sec: pov}

It is worth noting that the parity-odd vertex exists in case if one introduces another PM field, say $\vf'_{\m\n}$. Then, the corresponding vertex is
\begin{align}
\cV_{\vf\vf\vf'}=\vf^{\m\n}J'_{\m\n}\,,\quad J'_{\m\n}=F_{(\m}{}^{\r,\s}\tilde F'_{\n)\r,\s}-\frac14 \bar g_{\m\n} F_{\l\r,\s}\tilde F'^{\l\r,\s}\,,
\end{align}
where $F'_{\m\n,\r}=2\nabla_{[\m}\vf'_{\n]\r}$ is the curvature of the new PM field, and $\tilde{F}'_{\m\n,\r}=\frac12 \e_{\m\n\l\s}F'^{\l\s,}{}_{\r}$ is its dual.
The gauge transformations for the fields involved are given as
\begin{align}
\d_{\bar \a}^{\sst [1]} \vf_{\m\n}= \partial^\r \bar\a\, \tilde{F}'_{\r(\m,\n)}\,,\quad \d_{\bar \a}^{\sst [1]} \vf'_{\m\n}= -\partial^\r \bar \a\, \tilde{F}_{\r(\m,\n)}\,,\quad \d_{\a'}^{\sst [1]}\vf_{\m\n}=0=\d_{\a'}^{\sst [1]}\vf'_{\m\n}\,.
\end{align}
When identifying the two fields, their gauge transformations sum up to zero, which indicates the absence of the corresponding vertex for one field $\vf_{\m\n}$. Interestingly, when the two transformations have opposite sign, the corresponding commutator now gives an extra sign difference, which indicates that the extra field cannot help to close the commutator unless it is a ghost.

\section{PM Coupling to matter}
\label{sec: scalar coupling}

It is instructive to see how matter can couple to PM field. We can draw several conclusions from simple scalar coupling. First of all, we can notice easily that  if there is a scalar coupling to PM field, then the closure of the gauge algebra requires massless spin two in the game.
This can be easily seen in the following way. If there is a coupling of the scalar to PM, the scalar field should transform with the PM parameter. The most general transformation takes the form:
\be
\d^{\sst [1]}_{\a}\phi=b_1\, \partial^\r\a\,\partial_\r\,\phi+b_2\, \a\, \phi\,,
\ee
with the commutator of two transformations giving
\be
(\d_{\a_2}^{\sst [1]}\d_{\a_1}^{\sst [1]}-\d_{\a_1}^{\sst [1]}\d_{\a_2}^{\sst [1]})\,\phi\sim
\frac{2\,\L}{(D-1)(D-2)}\,b_1^2\,(\a_1\,\partial^\r\a_2-\a_2\,\partial^\r\a_1)\,\partial_\r\,\phi\,,
\ee
which is a diffeomorphism with the parameter,
\be
[\a_1\,,\a_2]^\r=\frac{2\,\L}{(D-1)(D-2)}\,b_1^2\,(\a_1\,\partial^\r\a_2-\a_2\,\partial^\r\a_1)\,.
\ee
The symbol $\sim$ stands for equivalence modulo terms that vanish for Killing parameters, or equivalently, that can be compensated by $[\d^{\sst [0]},\d^{\sst [2]}]$.

The scalar coupling has additional unexpected property: 
 there is non-trivial PM coupling only for the scalar with mass $m^2=\frac{D\,\L}{2\,(D-1)}$,
 which is nothing but the conformal scalar.
To illustrate this, let us write the most general spin-two current, bilinear in the scalar field:
\be
J_{\m\n}=c_1 \partial_\m\phi\,\partial_\n\phi
			+c_2 g_{\m\n}\partial_\r\phi\,\partial^\r\phi+c_3 g_{\m\n} \phi^2\,.
\ee
The partial conservation condition,
\be
\nabla^\m\nabla^\n J_{\m\n}+\frac{2\,\L}{(D-1)(D-2)}g^{\m\n}J_{\m\n}=0\,,
\ee
implies a system of three equations on parameters $c_i\,, \,i=1,2,3$ and the mass $m$ of the field $\phi$\,. In particular, this system gives
\begin{align}
c_2=-\frac12\, c_1\,,\quad c_3=-\frac12\, c_1\, (m^2-\frac{\L}{D-1})\,,\quad
(m^2-\frac{D\,\L}{2\,(D-1)})\,c_1=0\,.
\end{align}
Note that $c_1=0$ corresponds to a trivial interaction since the cubic vertex $\varphi^{\m\n}J_{\m\n}$ vanishes in that case. A non-trivial interaction exists only for $m^2=\frac{D\,\L}{2\,(D-1)}$.
In that case, there is a unique partially conserved current, up to an overall coefficient,
\be
J_{\m\n}=\partial_\m\phi\,\partial_\n\phi-\frac12 g_{\m\n}\,\partial_\r\phi\,\partial^\r\phi-\frac{(D-2)\,\L}{4\,(D-1)}g_{\m\n}\,\phi^2\,,
\ee
A conserved current (stress-energy tensor) for a scalar field with mass $m$ is given by
\be
T_{\m\n}=\partial_\m\phi\,\partial_\n\phi-\frac12 g_{\m\n}\,\partial_\r\phi\,\partial^\r\phi-\frac{m^2}{2}g_{\m\n}\,\phi^2\,,
\ee
for any value of $m^2$, whereas $J_{\m\n}$ is partially-conserved only for $m^2=\frac{D\,\L}{2\,(D-1)}$.  In fact, this is a particular case of a generic interaction for two scalars with different mass values and a PM field. As in the case of massive spin two fields, discussed in this paper, for massive scalar fields also there is a cubic interaction with a PM field only if the two massive scalars with masses,
\begin{align}
m^2(\m)=\frac{2\,\L}{(D-1)(D-2)}\,(\m+2)\, (D-3-\m)\,,
\end{align}
satisfy $\m_2=\m_1\pm 1$. In particular, the scalar with $\m_0=\frac{D}{2}-3$, can have an interaction with itself and a PM field, due to the fact that $m^2(\m_0)=m^2(\m_0+1)$. Note that this value of the scalar mass coincides with that of the conformal scalar.

We would like to note here that for massless spin one field the conserved current is also partially conserved in four dimensions only:
\be
J_{\m\n}=T_{\m\n}=F_{(\m}^{\quad\r}F_{\n)\r}-\frac14\, g_{\m\n}\, F_{\r\s}F^{\r\s}\,,
\ee
which is related to the fact that the Maxwell field is conformal in four dimensions. It is an easy exercise to show that there is no other value of mass for a vector field, for which one can couple Proca field to a PM spin two field in four dimensions. In arbitrary dimensions, coupling of the single Proca field \eqref{FreeProca} to PM field is given through a current,
\begin{align}
J_{\m\n}=F_{(\m}{}^{\r}F_{\n)\r}-\frac14\,g_{\m\n}\,F_{\r\s}F^{\r\s}-\frac{(D-4)^2\,\L}{2(D-1)(D-2)}(A_{(\m}A_{\n)}-\frac12\,g_{\m\n}\,A^{\r}A_{\r})\,,
\end{align}
that is partially conserved only for $m^2=\frac{(D-4)\,\L}{2(D-1)}$, which is again the mass value of the conformal Proca field, which coincides with the massless gauge field only for $D=4$.

It is therefore expected that a complete non-linear theory of a PM field will define a
specific spectrum of matter that can couple to it, giving a special role to conformal mass values.

\bibliography{PM}
\bibliographystyle{JHEP}

\end{document}